\documentclass[12pt]{extarticle}
\usepackage{lscape}
\usepackage[table,xcdraw]{xcolor}
\usepackage[english]{babel}
\usepackage[flushleft]{threeparttable}
\usepackage{tabularx} 
\usepackage{endnotes}
\usepackage{booktabs}
\usepackage{subcaption}
\usepackage{lmodern}
\usepackage{adjustbox}
\usepackage{chngcntr}
\usepackage[letterpaper,top=2cm,bottom=2cm,left=2.5cm,right=2.5cm,marginparwidth=1.6cm]{geometry}
\usepackage{float}
\usepackage{graphicx}
\usepackage[font=normalsize,labelfont=bf]{caption}
\makeatletter

\usepackage[utf8]{inputenc}
\usepackage[shortlabels]{enumitem}
\usepackage{amsmath, amssymb, verbatim, color, tikzsymbols, tabularx, appendix, booktabs, csquotes}
\usepackage[style=authoryear, citetracker=true, maxcitenames=2, hyperref=true, url=true, isbn=false, doi=true, natbib=true, 
backend=bibtex,
bibstyle=authoryear, maxbibnames=30, sorting=nyt, uniquename=full, uniquelist=false]{biblatex}
\usepackage[colorlinks=true, allcolors=blue]{hyperref}
\usepackage{setspace}
\usepackage[section]{placeins}
\usepackage{ragged2e}
\usepackage{makecell}
\setlist[description]{%
font={\bfseries\rmfamily}
}
\usepackage[affil-it]{authblk}

\newcolumntype{L}[1]{>{\raggedright\let\newline\\\arraybackslash\hspace{0pt}}m{#1}}
\newcolumntype{C}[1]{>{\centering\let\newline\\\arraybackslash\hspace{0pt}}m{#1}}
\newcolumntype{R}[1]{>{\raggedleft\let\newline\\\arraybackslash\hspace{0pt}}m{#1}}

\title{\Large The evolving boundary of green technology}
\author[1,2]{\normalsize Nicolò Barbieri
}
\author[3]{\normalsize Kerstin H\"otte}
\author[4,5]{\normalsize Peter Persoon}
\affil[1]{\normalsize Dept. of Economics and Management, University of Ferrara, Italy}
\affil[2]{\normalsize SEEDS Sustainability, Environmental Economics and Dynamics Studies}
\affil[3]{\normalsize Kedge Business School, Paris}
\affil[4]{\normalsize Wageningen University and Research}
\affil[5]{\normalsize Oxford Martin Programme on Technological and Economic Change, University of Oxford}
\date{}
\begin{document}

\maketitle

\begin{abstract} %
Green patent indicators are widely used to track technological progress, assess climate and innovation policies, and identify emerging technological opportunities. Yet these indicators depend on classification systems that evolve over time and shape which inventions become visible. 
Drawing on a sociological perspective, we interpret patent classifications as information infrastructures and examine three mechanisms that shape the visibility of green inventions: (1) modification of category boundaries through reclassification, (2) broadening of institutional coverage through set expansion, and (3) narrowing of scope through user-defined selection criteria.  
Using patent families filed between 1980 and 2016, we merge the 2019 and 2023 PATSTAT releases to isolate their effects.  Reclassification and set expansion increase the observed population of green patents by about 9\% and 10\%, respectively, while commonly used filters exclude between 42\% and 93\%. 
These effects are uneven and particularly pronounced for recent inventions, rapidly growing fields, and patents associated with Asian offices. Patents made newly visible through reclassification and set expansion are more novel and more often cited, while being slightly more specialized in narrow technological fields. 
Measurement choices can therefore change conclusions about the pace, direction, maturity, geography, and nature of green innovation. Observed green patent trends reflect not only inventive activity but also the evolving infrastructure through which green technology is made visible. Evaluations of policy effects, technological leadership, emerging opportunities, and progress in the green transition are therefore sensitive to classification vintages, institutional coverage, and filtering choices.

\end{abstract}

\vspace{0.2cm}

\vspace{0.2cm}

\newpage

\section{Introduction}
\label{sec:intro}
Accelerating the development and diffusion of green technologies is essential to achieve long-term climate goals \citep{ipcc2023report}.  
To govern this process, policymakers, firms, investors, and researchers need indicators that make technological change visible, comparable, and actionable. 
Patent-based indicators are widely used for this purpose because they allow to track the rate and direction of inventive activity, compare technological trajectories, assess policy inducement effects, and identify emerging areas of technological opportunity.\footnote{See e.g., \citet{popp2019complete} for a review of these studies}
These indicators do not merely describe green technological change; they help organize how such change is monitored and interpreted.

Recent patent-based evidence revealed ambiguous accounts of the recent trajectory of green technology. Several studies report a slowdown in green inventions during the 2010s, which has been a reason for concern in the face of accelerating climate change. The slowdown has been linked to falling fossil-fuel prices, low $CO_2$ prices, the financial crisis, and the increasing maturity of some renewable technologies \citep{popp2020innovation, aghion2022financial, martin2022private, probst2021global, iea2021drop}. 
At the same time, other recent studies indicate continued growth and, in some estimates, a subsequent rebound in green patenting during the 2010s \citep{wang_fueling_2024, wipo2026mapping}. 

This ambiguity raises a fundamental question: how do changes in the information infrastructure used to identify green technologies shape observed trends in green inventive activity?

We approach this question by acknowledging that the definition of green technology is neither undisputed nor static. Its boundaries evolve with scientific knowledge, policy priorities, environmental standards, and changing assessments of technologies within broader socio-technical systems \citep{golub2011defining, claro2007trade, cisneros2023defining}. Because sustainability transitions involve the co-evolution of technologies, institutions, infrastructures, markets, and practices \citep{geels2002technological, markard2012sustainability}, the climate relevance of particular technologies may change, affecting which inventions, technological trajectories, and actors are recognized as part of the green technology domain. The boundary of green technology is therefore not only a technical matter of definition; it is also part of ``how'' transition pathways are made legible and governable and of ``who'' participates in the process of determining ``what'' is green.

Patent classification systems are central to this process because they provide an information infrastructure for organizing technological knowledge encoded in measurable innovation outputs. They reduce search costs, facilitate comparability, and allow users to identify inventions that may be relevant for research, policy evaluation, technology strategy, and targeted support \citep{bowker2000sorting, hascic2015measuring}. At the same time, classification systems are not neutral representations of technological reality. They embed conventions, institutional judgments, and historically accumulated patterns of participation that shape what becomes visible, measurable, and comparable. 

In this paper, we examine how the evolution of the information infrastructure underlying the identification of climate change mitigation and adaptation technologies affects the measurement of green innovation. Specifically, we focus on the Y02 tagging scheme of the Cooperative Patent Classification (CPC) system, which was originally developed to make climate-related technologies searchable within patent databases and has since become the dominant tool for identifying green inventions, tracking green innovation trends, and evaluating the effects of environmental and innovation policies \citep[e.g.][]{barbieri2023green, persoon2020science, probst2021global, hotte2021rise, hotte2022knowledge}. 

We identify three analytically distinct mechanisms of infrastructural change, each operating at a different layer of the measurement system. First, \emph{reclassification} captures changes in category boundaries: as Y02 categories are introduced, modified, reorganized, or removed, patents may retrospectively enter or leave the green domain. Because the greenness of technologies depends on policy objectives, environmental assessment criteria, social legitimacy, and their role within evolving socio-technical systems, reclassification may represent delayed recognition of technological trajectories whose climate relevance becomes clearer over time \citep{geels2002technological, markard2012sustainability, golub2011defining}. 

Second, \emph{set expansion} captures changes in institutional coverage: as CPC coverage broadens across patent offices, inventions already present in the global green technological landscape become newly legible through the measurement system. Because sustainability transitions are spatially uneven and shaped by regionally specific capabilities, industrial structures, and policy priorities \citep{truffer2015geography, coenen2012geography}, delayed coverage can affect conclusions about the locus and timing of green technological leadership. 

Third, \emph{filtering} captures selective use of the infrastructure: researchers, policymakers, and analysts apply criteria such as citations, family size, or triadic status to retain patents considered economically or technologically relevant. Although often reasonable, these filters imply subjective judgments and may privilege established, internationally embedded, or path-continuing inventions while excluding more local, emerging, or less institutionally connected ones \citep{higham2021patent, squicciarini2013measuring}. 

Together, these mechanisms do not merely affect patent statistics; they reshape the information infrastructure through which green technological change becomes visible. By determining which inventions are included, excluded, or reclassified, they influence the identification of emerging technological opportunities, the measurement of technological leadership, and the policy decisions that guide resource allocation and innovation support.

The present paper asks how the three mechanisms identified above influence the measurement and broader understanding of green technological change from three angles. First, the mechanisms may alter assessments of technological leadership and catching-up. Technological leadership and specialization are the results of cumulative learning, diversification, and technological change \citep{malerba2011learning,  corrocher2024green, perruchas2020specialisation}. If countries specialize in technologies that are classified as green only retrospectively, or if their patent offices enter the CPC infrastructure with a delay, changes in measured performance may partly reflect changes in visibility rather than changes in inventive activity. Filtering may reinforce this effect when selection criteria favor patents filed through internationally central offices or protected through large international families.

Second, the mechanisms may affect the observed technological opportunity space. Patent classifications are used to measure technological capabilities, relatedness, and diversification. Changing boundaries can therefore alter which technological trajectories appear established, adjacent, or emerging, and also influence the perceived opportunity space for smart specialization and industrial policy \citep{hidalgo2007product, neffke2011regions, montresor2020green, barbieri2023regional}. Technologies that do not fit existing categories may remain only partially visible until the classification system evolves. 

Third, the mechanisms may alter the observed nature of green innovation. If included and excluded patents differ systematically in their knowledge base, novelty or subsequent impact, classification vintages and filtering choices can change whether green technological change appears incremental or path-breaking, specialized or pervasive \citep{barbieri2020knowledge, dahlin2005radicalness}. Moreover, these metrics rely on classifications and references among patents observable within the infrastructure, making qualitative assessments of green technology sensitive to the infrastructure's evolution. 
While path-continuing innovations are more likely to fit existing measurement methods, path-breaking, general-purpose, or system-level inventions may be recognized only with delay, may initially fall outside established green categories, or may be weakly represented in dominant international patenting networks \citep{dahlin2005radicalness, bresnahan1995general}. 

Empirically, we study these mechanisms by constructing an original dataset that integrates the 2019 and 2023 releases of PATSTAT, the global patent repository maintained by the European Patent Office (EPO). We construct matched samples of DOCDB patent families with earliest filing years between 1980 and 2016. Comparing the two releases allows us to isolate green patent families affected by reclassifications, set expansion and filtering effects. We then examine how commonly used citation-, family-, and jurisdiction-based filters alter observed trends and interact with the other two mechanisms.

The results show that, from a quantitative perspective, reclassification and set expansion effects increase the number of observed green inventions by 9.2\% and 10.6\%, respectively, over the period 1980--2016, while filtering has an even larger quantitative impact, reducing the number of green patent families by between 42\% and 93\% relative to all green patent families retrieved in the dataset.

The effects are unevenly distributed and affect interpretations in three regards. On the one hand, these changes affect assessments of technological leadership and catching-up. Set expansion is concentrated in patent families filed at Asian offices, particularly those in Japan, China, and South Korea, indicating that their contributions were underrepresented in the earlier database release. Jurisdiction- and family-based filters also disproportionately exclude patents associated with these offices. Apparent differences in countries' green technological performance may therefore reflect not only inventive activity, but also the timing of CPC coverage and the criteria used to construct patent samples. 

In addition, these mechanisms affect the relevance of some green technologies which may distort the observed technological opportunity space. Reclassification and set expansion are unevenly distributed across technological domains, with particularly large effects in rapidly growing fields such as batteries and clean production. Analyses based on earlier classification vintages may therefore underestimate the scale and momentum of some emerging trajectories and misrepresent their position relative to more established technologies.

Third, classification vintages and filtering choices may affect assessments of the nature of green innovation, although the differences observed across groups are generally modest. Patents brought into view through reclassification and set expansion are, on average, somewhat more technologically specialized, more novel, and more frequently cited than comparable green patents. The family size filter excludes inventions that tend to be slightly less original and less novel, but more general, and to receive fewer five-year forward citations. These patterns suggest that measurement choices can influence how the green technology frontier is characterized--as relatively broad or specialized, cumulative or novel, and more or less technologically influential--while the magnitude of these effects remains limited.\footnote{The comparisons drawn here are exclusively internal to the population of green patents. They concern differences among green technologies and must not be interpreted as comparisons between green and non-green patents.}

The findings show that green patent statistics are shaped by an evolving information infrastructure, not only by underlying inventive activity \citep{bowker2000sorting, nelson2014innovation}. 
For researchers and policymakers, reclassification, set expansion, and filtering may alter assessments of technological leadership, emerging opportunities, and the nature of green innovation. For firms and investors, reliance on a single classification vintage or a restrictive patent sample may obscure own and competitors' capabilities in late-covered jurisdictions and emerging technological domains. These measurement choices can therefore affect policy evaluation, country benchmarking, technology search, and the allocation of attention and resources across competing transition pathways.

The remainder of the article is organized as follows. Section~\ref{sec:lit_rev} reviews the literature on evolving classification infrastructures and their implications for technological leadership, opportunity identification, and the observed nature of innovation. Section~\ref{sec:conceptual_framework} develops the conceptual framework and introduces reclassification, set expansion, and filtering. Section~\ref{sec:data_methods} describes the construction of the multi-release PATSTAT dataset and the empirical strategy. Section~\ref{sec:findings} presents the results, and Section~\ref{sec:discussion} discusses their implications for research, policy evaluation, and technology strategy.

\FloatBarrier

\section{Literature review}
\label{sec:lit_rev}

\subsection{Green technology classifications as evolving information infrastructures}
\label{subsec:lt_info_infrastructure}

Categories make heterogeneous objects measurable and comparable by specifying which similarities and differences matter. A classification system organizes such categories into a structured scheme and establishes rules for assigning objects to category labels. When embedded in databases, organizations, and shared practices, a classification system becomes part of an information infrastructure: a socio-technical arrangement through which complex information is collected, standardized, sorted, exchanged, and used. Such infrastructures enable coordination, but the boundaries and assignment rules they encode also shape what becomes visible and what remains outside the measurement system \citep{bowker2000sorting, star1999ethnography, lamont2002study}.

The boundaries of the green category are continuously redefined through scientific knowledge, policy agendas, environmental standards, industrial strategies, and changing interpretations of how particular technologies contribute to the fight against climate change. Definitions of what counts as green therefore differ across time, institutional settings and fields of economic activity, such as finance, product markets and innovation \citep{golub2011defining, claro2007trade, cisneros2023defining}. Such dynamic nature is the result of the co-evolution of technologies, institutions, markets, infrastructures, and user practices \citep{geels2002technological, markard2012sustainability}, requiring classifications to accommodate both new inventions and changing interpretations of existing ones.

Patent classification systems translate technological domains into operational categories. They organize patent documents by technical subject, support prior-art search, and allow users to identify narrow technological fields through a hierarchical infrastructure \citep{simmons2014categorizing, hascic2015measuring}. In green innovation research, the Y02 classes of the CPC system have become a widely used tool to identify inventions related to climate-change mitigation and adaptation \citep{angelucci2018supporting, probst2021global, hotte2022knowledge, barbieri2023green}. Indicators based on these classes therefore depend on the category definitions and institutional coverage of a given version of the classification infrastructure.

From an infrastructure perspective, three mechanisms may change the visibility of inventions. Revisions to category definitions (reclassification), changes in the institutional coverage (set expansion), and choices made by infrastructure users (filtering) can alter which inventions, technologies, and geographies are visible in empirical data. The following subsections examine the background of how these mechanisms affect assessments of technological leadership and geography, the identification of technological opportunities and diversification pathways, and the observed nature of green innovation.

\subsection{Technological leadership and the changing geography of green innovation}
\label{subsec:lt_leadership}

Technological leadership in a given domain reflects accumulated knowledge, capabilities, and prior specialization. Because these capabilities develop cumulatively, leadership and specialization often persist along path-dependent technological trajectories \citep{dosi1982technological, boschma2017relatedness}. It may nevertheless shift during technology transitions when changes in technology, demand, or institutions create windows of opportunity for latecomer countries and firms to challenge incumbents and establish competitive advantages along emerging technological trajectories \citep{lee2017catch}.

The literature on technological catch-up shows that leadership frequently changes through processes of capability accumulation, knowledge acquisition, and strategic specialization \citep{malerba2011learning, lee2017catch}. Latecomer economies may initially rely on imitation and knowledge transfer but can progressively develop indigenous innovative capabilities and eventually compete at the technological frontier. In some cases, technological discontinuities create opportunities for leapfrogging, allowing new entrants to bypass established trajectories and overtake incumbent leaders \citep{lee2017catch}. Historical evidence shows that leadership transitions have repeatedly occurred across sectors and countries, driven by the interaction of technological opportunities, institutional arrangements, and industrial capabilities. A notable example is the technological catch-up of Korea, Taiwan, and China, which have converged toward the technological frontier at different speeds and across different dimensions over the last decades \citep{kwon2017international}. 

These dynamics are salient in green technologies, as the sustainability transition is generating new technological opportunities while reshaping existing competitive advantages. 
Countries tend to diversify into green domains related to their existing knowledge bases, producing persistent but differentiated patterns of specialization \citep{perruchas2020specialisation, santoalha2021diversifying}. 
International knowledge spillovers can support diversification by follower and catching-up countries, while leaders tend to enter more complex green technologies at earlier stages of development \citep{corrocher2024green}. 
As a result, leadership varies across technological domains, with different countries occupying frontier positions in renewable energy, batteries, energy efficiency, smart grids, electric mobility, and adaptation technologies. 

Our paper contributes to the literature on green technology leadership by showing that evolving information infrastructures may affect the measurement of inventive activity and countries’ relative performance. 

\subsection{Technological opportunities and diversification}
\label{subsec:lt_opportunities}

Technological diversification of countries, regions, and firms is path-dependent. Regions are more likely to enter technological fields related to capabilities already present in their knowledge bases, while entry into unrelated fields generally requires capabilities that are more difficult to access or develop \citep{neffke2011regions, boschma2017relatedness}. The set of technological fields that appears accessible for future specialization and diversification from the existing capability base constitutes an observed technological opportunity space and  promising avenues for future innovation.

Green technological diversification follows similar patterns. Studies at the country and regional levels show that entry into green technological fields depends on related capabilities, the maturity of the technology, policy support, and complementarities between green and non-green knowledge bases \citep{montresor2020green, perruchas2020specialisation, santoalha2021diversifying, barbieri2023green, barbieri2023regional}. Patent-based measures of technological relatedness are therefore used not only to describe existing specialization but also to identify fields into which countries or regions may be able to diversify.

These evaluations depend on classification systems and the database used to measure inventive activity. Patent categories define the technological units from which specialization, relatedness, and diversification potential are inferred. When categories are revised, previously uncovered patent offices become visible, or analytical filters remove parts of the green technological landscape, the observed distribution and proximity of technological capabilities may also change. The vintage of the information infrastructure and sample construction choices can therefore alter which green fields appear established, related, or emerging, even when the underlying inventive activities have not changed. 

In this paper, we examine whether these three mechanisms affect the share of different technological fields within the green domain and, in doing so, influence the shape of opportunity spaces.

\subsection{The observed nature of green technology}
\label{subsec:nature_green_innovation}

The nature of green inventions can be assessed from complementary ex-ante and ex-post perspectives. 
Ex-ante perspectives evaluate the technological search process and recombination that led to an invention, including the breadth of the technological fields it spans, the diversity of the prior technological and scientific knowledge on which it builds, and the novelty of the combinations it introduces.
Ex-post perspectives focus on the invention's influence on subsequent innovation, including both the intensity and technological breadth of that influence \citep{fleming2001recombinant, verhoeven2016measuring, barbieri2020knowledge}. 

Ex-ante and ex-post measures provide complementary means to identify path-continuing and path-breaking technological change. Path-continuing inventions advance established trajectories through modifications and recombinations within an existing technological paradigm, whereas path-breaking inventions depart from established search directions and contribute to the emergence of new trajectories \citep{dosi1982technological}. 
Unusual knowledge recombinations may indicate path-breaking potential, but realized path-breaking impact is only established ex post, when an invention redirects or substantially influences subsequent technological development \citep{dahlin2005radicalness, verhoeven2016measuring}. Ex-ante novelty is therefore an indicator of path-breaking potential but provides no evidence of realized technological impact \citep{fleming2001recombinant}. 

These distinctions are particularly relevant for green innovation, which encompasses both incremental improvements along established trajectories and discontinuous changes in technological configurations and knowledge bases. Green inventions may improve mature technologies, combine knowledge from previously separate domains, or open new pathways. Empirical studies suggest that green patents have distinctive search and impact profiles. \citet{barbieri2020knowledge} find that green patents combine more technological components, draw on more diverse knowledge sources, and exhibit more novel recombinations than comparable non-green patents. They also generate larger and more pervasive impacts on subsequent invention. \citet{fusillo2023green} similarly documents greater technological diversity in both the knowledge-search and knowledge-output phases of green invention. For energy and transport technologies, \citet{dechezlepretre2014knowledge} find larger citation-based spillovers from clean than from dirty patents. 

This evidence suggests that green inventions differ from other non-green inventions both in how knowledge is recombined and in how their influence diffuses across subsequent technological development. 
Since green inventions often emerge from diversified and cognitively distant knowledge domains and their impact spans across fields, their identification and assessment may be particularly sensitive to the classification infrastructure through which technological recombination and impact are observed. 

Revisions to category boundaries, changes in patent office coverage, and analytical filters may include or exclude patents with systematically different search and impact profiles. Classification schemes, data vintages, and sample construction choices may contribute to differences in the observed distributions of technological scope, originality, novelty, forward citations, and generality.
The next section develops a conceptual framework explaining how changes in classification, institutional coverage, and filtering shape the observed boundary of green technology. 

\section{Conceptual framework}
\label{sec:conceptual_framework}
Conceptualizing the observed green technology domain as part of an evolving information infrastructure, we distinguish three mechanisms (i.e., reclassification, set expansion, and filtering) that operate at different layers of the information-retrieval process, shape this observed domain, and guide our empirical analysis. 

\subsection{Reclassification and changing technological meaning}
\label{subsec:reclassification_theory}

The first mechanism, i.e., reclassification, concerns revisions to the CPC scheme that change the categories assigned to existing and future patent documents. As technologies develop, classifications are updated to accommodate new or growing technological fields and maintain search efficiency in an evolving landscape \citep{lafond2019long, epo_cpc_monitoring_2023}. 

Categorization involves a trade-off between coarse categories, which simplify the search of information across many objects, and fine-grained categories, which distinguish more precisely among heterogeneous objects \citep{mohlin2014optimal}. 
The clarity of category boundaries can also affect attention: patents assigned to more sharply delineated technological classes receive more citations than otherwise comparable inventions \citep{kovacs2021categories}. 
As technological knowledge develops, existing categories may therefore become too broad, ambiguous, or poorly aligned with emerging fields. Patent classification systems respond by creating, splitting, merging, or deleting categories and by retrospectively reassigning existing patents \citep{lafond2019long}. 

Within the CPC, such changes are implemented through revision and maintenance projects. 
The process can be summarized in three steps. First, a need for revision is identified and a revision or maintenance project is initiated. Second, classification specialists revise the scheme and, where necessary, its definitions, concordances, and assignment rules. The proposed changes are published before their entry into force. Finally, once the revised scheme enters into force, it is used to classify new documents and, where required, to reassign documents to the revised categories.

Because retrospective reassignment may take time, old and new categories can coexist temporarily until the  reclassification is completed. 
Revision of category boundaries is an institutional process, whereas the assignment and reassignment of documents may be supported and accelerated by automation. At the EPO, the allocation of Y02 and Y04 symbols to incoming documents was fully automated in 2022, and related artificial intelligence tools provide suggestions for the reclassification of existing patent and non-patent literature \citep{epo_ai_y02_2022}. Automation can therefore aid in the implementation, but decisions on the scope and structure of the category revisions remain institutionally governed and involve human judgment. 

Since its inception in 2013, the Y02 scheme has been expanded and reorganized repeatedly. In its first version, the scheme included subclasses covering buildings, greenhouse-gas capture, energy, and transport (Y02B, Y02C, Y02E, and Y02T). Waste and wastewater treatment technologies (Y02W) were added in May 2015, production technologies (Y02P) in November 2015, and adaptation technologies and ICT (Y02A and Y02D) in January 2018 \citep{veefkind2012new, angelucci2018supporting}. 
Existing subclasses have also been revised through the deletion, introduction, consolidation, and reorganization of groups \citep{epo_uspto_cpc_archive}. 
Table~\ref{tab:history_recla} summarizes the main scheme changes between 2015 and 2020.

From an information infrastructure perspective, reclassification changes the operational boundary of green technology by introducing new categories or revising existing ones. Existing inventions may consequently enter or leave the green domain, so increases in observed green patenting can reflect retrospective reassignment as well as new inventive activity. 
Because such revisions accumulate, classification vintages may represent the same body of inventions differently \citep{lafond2019long}. Reclassification is more frequent among recent patents \citep{persoon2026reclassification}, making database vintage particularly consequential for emerging technological trajectories. Research on innovation indicators similarly shows that alternative labels and indexing systems can yield different estimates of the timing, scale, and composition of emerging fields \citep{nelson2014innovation}. 

In our setting, reclassification therefore captures changing, and sometimes delayed, recognition of inventions whose relevance to the green domain becomes clearer over time. Table~\ref{tab:conceptual_framework} summarizes the implications for observed technological leadership, opportunity spaces, and the nature of green innovation.

\subsection{Set expansion, institutional coverage, and visibility}
\label{subsec:set_expansion_theory}

The second mechanism, set expansion, changes which inventions are covered by the Y02-scheme, and therefore visible members of the green category. 
Coverage may expand as patent offices introduce or extend CPC classification, classify earlier documents retrospectively, or make additional family-level classification information available \citep{DEGROOTE2018S78}. 

From an information-infrastructure perspective, set expansion concerns institutional connectedness to the infrastructure. 
Classifications and standards create shared categories that enable coordination, but they also reflect the actors, practices, and knowledge bases that were present when the categories were designed and implemented \citep{bowker2000sorting, star1999ethnography}. A jurisdiction that is weakly connected to the CPC infrastructure may therefore be present in the underlying technological landscape while remaining less visible in the indicators built from that infrastructure. 
In other context, visibility is partly shaped by network embeddedness: actors with stronger connections to central institutions, markets, and knowledge networks are more likely to be recognized, compared, and evaluated through dominant categories \citep{zuckerman1999categorical, uzzi1997social}. 
Here, embeddedness can be considered the institutional integration into the CPC infrastructure through participating patent offices, databases, standards, and classification practices.
The institutional network through which the CPC is implemented shapes which parts of the global innovation landscape are visible in patent-based indicators. 

Patents may already exist in the global patent system but remain absent from Y02 indicators because the relevant documents are not yet covered by CPC classifications. 
The expansion of coverage makes existing inventive activity newly observable.

Set expansion can be geographically non-neutral because technological capabilities, patenting practices, and CPC coverage are unevenly distributed across jurisdictions. Earlier research found that patent families not covered by the CPC were disproportionately associated with the Japanese, Korean, and Chinese patent offices \citep{DEGROOTE2018S78}. If late-covered offices specialize in particular technological domains, expanded coverage may change not only the number of observed green inventions but also their apparent geographical and technological distribution \citep{truffer2015geography}. What appears as a sudden change in technological leadership may partly reflect the delayed visibility of existing inventions that become connected to the classificatory infrastructure.

These effects can extend beyond measured leadership, as summarized in Table~\ref{tab:conceptual_framework}. If late-covered jurisdictions have different technological specializations, set expansion may alter the observed opportunity space, and if lately visible patents have different search and impact profiles, it may also change the observed nature of green innovation. 

\subsection{Filtering, patent indicators, and selective recognition}
\label{subsec:filtering_theory}

The third mechanism is filtering, i.e., the application of user-defined criteria to select a group of patents already made visible through the classification infrastructure. Researchers often use forward citations, international family size, triadic status, or filings at selected patent offices to retain inventions expected to be technologically or economically relevant \citep{squicciarini2013measuring, higham2021patent}. Such filters can be useful for constructing samples suited to particular research questions, but they capture different perspectives on patent quality and value, which may vary across inventors, policymakers or the general public, and do not necessarily overlap \citep{higham2021patent}. Filtering does not change whether an invention is assigned to a Y02 category, but it determines which Y02-classified inventions enter the analytical sample. 

Because patent indicators capture distinct dimensions of technological and economic relevance, different filters need not retain the same inventions \citep{higham2021patent}. Filter choice is therefore an analytical decision rather than a neutral means of removing low-quality patents. 
Citation filters select inventions according to subsequent technological influence, but citation counts accumulate over time and vary with technological field, patent office procedures, and applicant and examiner practices \citep{jaffe2017citations, alcacer2006citations, criscuolo2008citations}. They are therefore not directly comparable measures of patent quality across jurisdictions. 
Family size and triadic filters select inventions protected across multiple jurisdictions and may indicate expected private value or international market scope. However, they also reflect applicants’ resources, commercial strategies, and choices about where to seek protection \citep{squicciarini2013measuring, dernis2015world}. Office-based filters similarly provide a common institutional setting but may favor domestic applicants and inventions pursued through established international filing routes \citep{criscuolo2006home}. As green inventive activity becomes more geographically dispersed, these filters may underrepresent locally oriented inventions and jurisdictions with different patenting practices \citep{kwon2017international, corrocher2024green}. 

From an information infrastructure perspective, filtering acts as a user-imposed eligibility rule. It can change both the size and composition of the observed green patent population, potentially underrepresenting recent inventions, locally oriented patents, and activity in jurisdictions with different filing or citation practices. Its implications extend across the three dimensions considered in this paper: filters may alter measured technological leadership, exclude trajectories from the observed opportunity space, and change the assessed search and impact profiles of green innovation (see Table~\ref{tab:conceptual_framework}).

\FloatBarrier

\begin{table}[htbp]
\centering
\caption{Mechanisms shaping the observed boundary of green technology and their implications}
\label{tab:conceptual_framework}
\footnotesize
\begin{tabular}{p{2.6cm} p{4.1cm} p{4.1cm} p{4.1cm}}
\toprule
\textbf{Mechanism} 
& \textbf{Technological leadership} 
& \textbf{Technological opportunities} 
& \textbf{The nature of green technology} \\
\midrule

\textbf{Reclassification} 
& Countries specialized in technologies that are later reclassified may appear to gain or lose green technological performance after boundary changes. 
& Emerging, contested, or system-level trajectories become visible only when classifications evolve, reshaping the observed opportunity space. 
& Reclassified patents may differ in novelty, scope, generality, and citation impact, changing the perceived radicalness or maturity of some green technologies. \\

\midrule

\textbf{Set expansion} 
& Late-visible jurisdictions may appear weaker before CPC adoption and may seem to catch up suddenly once connected to the infrastructure. 
& Newly visible offices may specialize in different green technology domains, shifting the observed structure of diversification opportunities. 
& Newly visible patents may differ in technological profile because patenting practices, office procedures, and sectoral specializations vary across jurisdictions. \\

\midrule

\textbf{Filtering} 
& Filters may privilege internationally embedded actors and incumbent patenting systems, thereby affecting measured country performance. 
& Local, emerging, or less internationally extended trajectories may be excluded from maps of green technological opportunities. 
& Filtered-out patents may differ in novelty, generality, and impact, affecting conclusions about the quality and spillover potential of specific green technologies. \\

\bottomrule
\end{tabular}
\end{table}

\FloatBarrier

\section{Data and methods}
\label{sec:data_methods}

\subsection{Data sources and sample}

We combine the Spring 2019 and Spring 2023 editions of PATSTAT Global, the EPO's worldwide patent-statistics database. Each edition provides a snapshot of the patent documents, bibliographic records, family relationships, and classification assignments available at that date.

Our unit of analysis is the DOCDB patent family to avoid double counting of the same invention when protection is sought through several patent offices. We assign each family to its earliest filing year and collapse the CPC assignments at the family level. We define a patent family as green in a given PATSTAT edition if at least one family member carries at least one Y02 symbol in that edition. 

We retain patent families with earliest filing years between 1980 and 2016. The end year reduces right truncation in the Spring 2019 edition, since patent applications are generally published around 18 months after filing or, where priority is claimed, after the earliest priority date. 
The cutoff does not eliminate all vintage effects: documents may be incorporated with delays, earlier patents may be classified retrospectively, patent families may acquire additional members, and forward citations continue to accumulate. 

The analysis involves two distinct temporal dimensions. The earliest filing year describes when an invention entered the patent system, whereas the PATSTAT edition date describes the timing of database and classification version. 

\subsection{Identification of the three mechanisms}

\begin{figure}[h!]
    \centering
    \caption{Dataset construction using PATSTAT 2019 \& 2023}
    \label{fig:data_recla_expan}
    \begin{subfigure}{0.49\textwidth}
        \includegraphics[width=\linewidth]{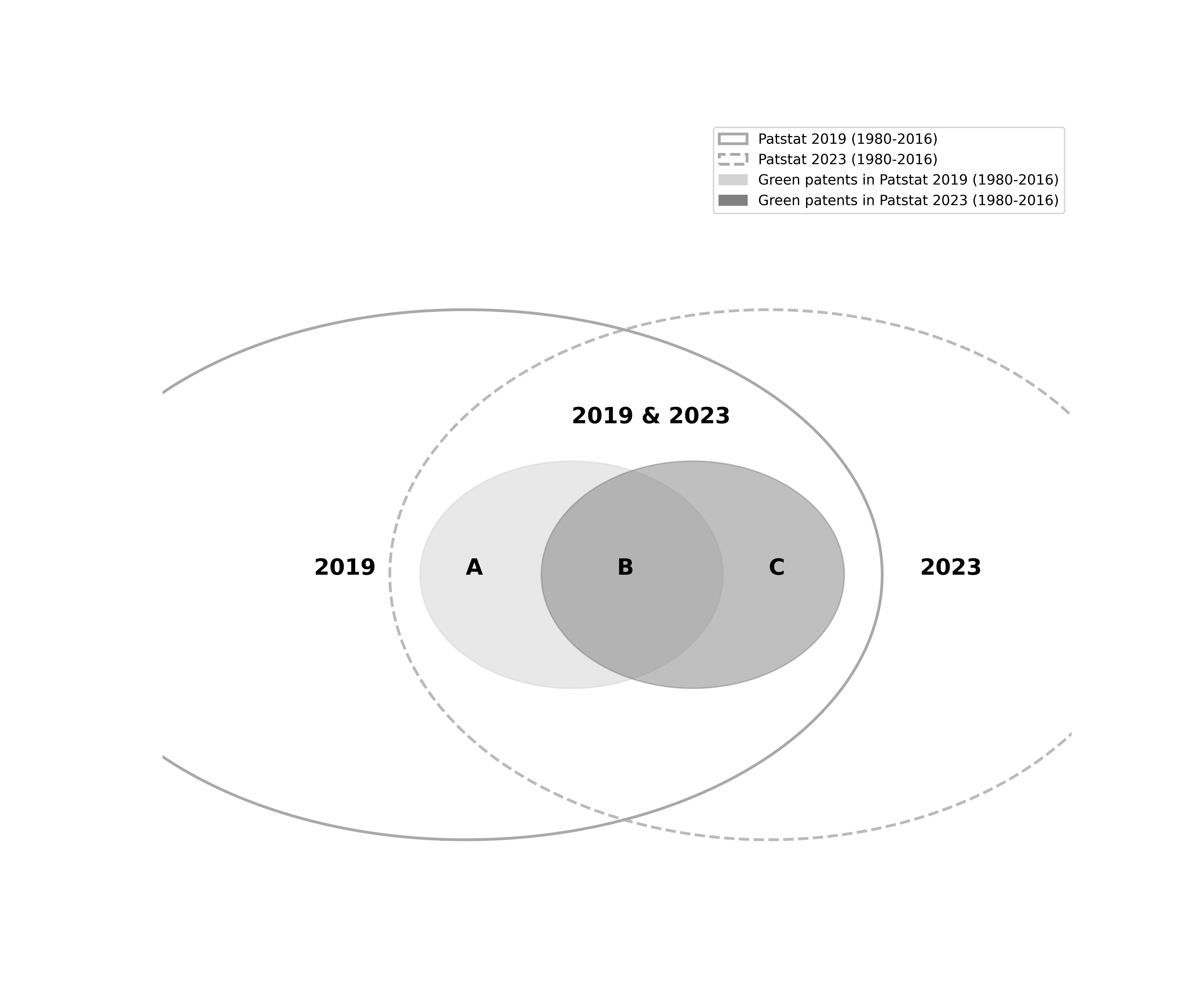}
        \caption{Reclassification data}
        \label{subfig:data_recla}
    \end{subfigure}
        \begin{subfigure}{0.49\textwidth}
        \includegraphics[width=\linewidth]{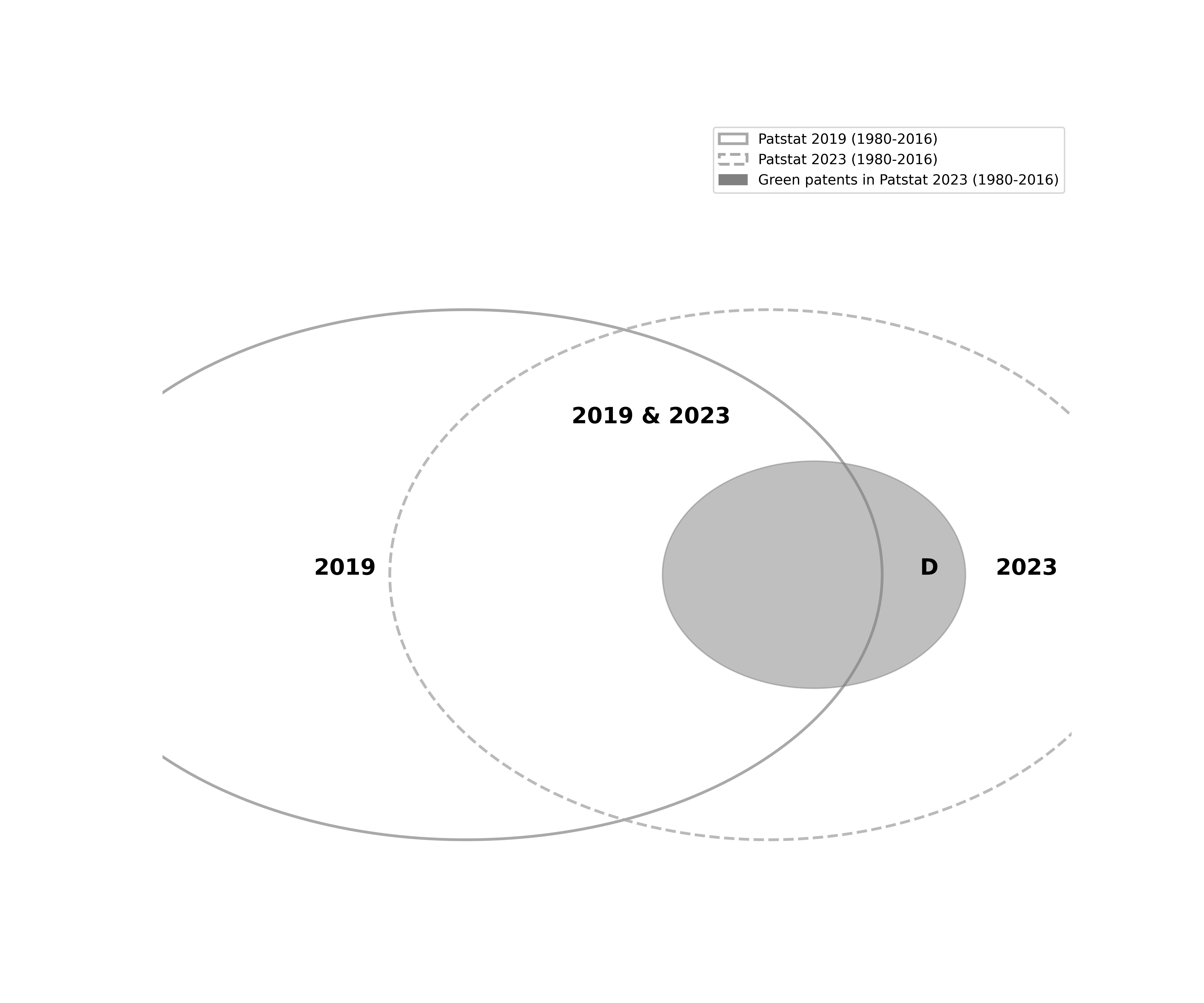}
        \caption{Set expansion data}
        \label{subfig:data_expa}
    \end{subfigure}

    \justifying \noindent \footnotesize

    Notes: The Venn diagrams illustrate how the datasets were constructed to analyze the reclassification and set expansion effects. Patent families affected by the reclassification effect (Figure~\ref{subfig:data_recla}) correspond to group C --that is, inventions not classified as green in the PATSTAT 2019 version but assigned at least one Y02 tag in the PATSTAT 2023 release. Patent families subject to the set expansion effect (Figure~\ref{subfig:data_expa}) are identified exclusively in the PATSTAT 2023 version (group D), although their priority year falls within the 1980–2016 period, they were not observable in the corresponding 2019 dataset construction.
    
\end{figure}

Comparing the two PATSTAT editions allows us to construct three analytical comparison samples corresponding to reclassification, set expansion, and filtering. Figure~\ref{fig:data_recla_expan} illustrates the construction of the first two. 

To identify reclassification, we restrict the sample to patent families observed in both PATSTAT editions. We then compare their Y02 assignments across versions. 
Three potential lists of green patents can be constructed as illustrated by the Venn diagram reported in Figure ~\ref{subfig:data_recla}: (A) patents that were green in 2019 and not in the 2023 PATSTAT edition; (B) patents that were green in both versions; and (C) patents that were non-green in the 2019 version but became green due to reclassification, which forms our reclassification dataset. A reclassification from non-green to green occurs when none of the patent family members carries a green CPC symbol in 2019, but one of the family members was reclassified as green in the 2023 release.

This family-level definition provides a lower-bound estimate. Changes to individual family members are not counted when they do not alter the green status of the family as a whole. We also exclude a negligibly small number of families that move from green in 2019 to non-green in 2023.

To identify set expansion, we hold the green classification definition constant using the Y02 assignments observed in the 2023 release (gray area in Figure \ref{subfig:data_expa}). We sample the patents classified as green in PATSTAT 2023 but absent from PATSTAT 2019 despite having an earliest filing year from 1980 to 2016 (area D in Figure \ref{subfig:data_expa}). These families constitute the set expansion sample. 

This sample captures the increase in the number of green patents that become visible through the expanded CPC coverage at patent offices, retrospective classification of earlier documents, delayed incorporation of records into PATSTAT, or updates to patent family information. Because these families are absent from the 2019 edition, we cannot observe how they would have been classified under the 2019 Y02 scheme. Set expansion should therefore not be interpreted exclusively as the effect of formal CPC adoption. 

To study filtering, we apply five alternative selection criteria separately to the samples of green patents observed in each PATSTAT edition. The family size filter retains families with applications in at least two patent jurisdictions. The citation filter retains families with at least one forward citation within five years after the earliest priority year. 
The EPO and USPTO filters restrict the samples to families containing at least one application filed at the respective office. The triadic filter retains families containing applications at the EPO, USPTO, and Japan Patent Office \citep{criscuolo2006home, squicciarini2013measuring}.

The indicators are measured separately in each database edition. A family may therefore fail a filter in the 2019 edition but satisfy it in 2023 because additional family members or citations have subsequently become observable. This temporal dependence is part of the filtering mechanism. 

\subsection{Assessing the nature of excluded inventions}
\label{subsec:patent_characteristics_methods}
We examine whether the three mechanisms may affect the observed nature of green technology. Unlike studies comparing green and non-green inventions \citep[e.g.,][]{barbieri2020knowledge}, we compare samples of green inventions defined by the three mechanisms with other green patent families always visible in the green technological domain.

We characterize affected inventions using complementary ex-ante and ex-post measures. The ex-ante measures include \emph{technological scope}, defined as the breadth of technological fields represented in the patent family. Empirically, it is measured as the total number of technology classification codes assigned to patents. In addition, \emph{originality}, defined as the diversity of technological fields represented in its cited prior art \citep{lerner1994importance, trajtenberg1997university, barbieri2020knowledge}. It is calculated as one minus the Herfindahl concentration index of the technological classes represented in a patent’s backward citations. Then, we measure \emph{combinatorial novelty} through technological combinations not previously observed in the patent record, using four-digit classifications to capture novel combinations across broad technological fields \citep{verhoeven2016measuring}. 
The ex-post measures capture subsequent technological influence. Forward citations are counted over five-year windows, while \emph{generality} measures the breadth of that influence across technological fields \citep{trajtenberg1990penny, jaffe2017citations}.\footnote{The index is measured in the same way as originality, but using forward citations instead of backward citations.}

For each mechanism $m \in \{\text{Reclassification, Set expansion, Filtering} \}$ and patent characteristic $I \in \{\text{Scope, Originality, Novelty, Generality, Citations}\}$), we estimate the following model:

\begin{equation}
\text{Pat}_{f}^{I} = \alpha + \beta \text{Affected}^m_{f} + \gamma \text{TechDom}_{f}
+ \delta \text{Time}_{f} + \lambda \text{PatOff}_{f} + \varepsilon_{f},
\label{eq:patent_characteristics}
\end{equation}

where $\text{Pat}_{f}^{I}$ denotes the value of patent characteristic $I$ for patent family $f$. 
$\text{Affected}^m_{f}$ is a binary indicator equal to one if patent family $f$ belongs to the exclusion sample generated by mechanism $m$, and zero otherwise. The coefficient of interest, $\beta$, captures whether affected patents differ systematically from those that remain visible in the baseline green patent sample. $\text{TechDom}_{f}$ denotes a set of dummy variables that capture the idiosyncratic features of green technological domains at 8-digit Y02 CPC code assigned to patent family $f$. $\text{Time}_{f}$ are five-year-period fixed effects based on the family's earliest filing year, controlling for temporal changes in technological opportunities and patenting activity. $\text{PatOff}_{f}$ is a set of patent office dummies identifying the jurisdiction(s) in which the patent family was filed, capturing institutional differences in patenting practices and classification procedures across offices. Finally, $\varepsilon_{f}$ is the idiosyncratic error term.

Because technological scope is a non-negative count variable, we estimate the scope specifications using Poisson regression. Originality, novelty, generality, and forward-citation specifications are estimated using ordinary least squares. All models report heteroskedasticity-robust standard errors.

\FloatBarrier
\section{Findings}
\label{sec:findings}

\subsection{Baseline scenario}
\label{subsec:baseline_trends}
We begin by comparing unfiltered counts of green patent families across the two PATSTAT editions. In each release, the baseline sample consists of all green DOCDB families, without imposing any filter. This comparison documents the overall difference between the green patent populations observable in 2019 and 2023. 

\begin{figure}
    \centering
        \caption{Total number of green patent families in PATSTAT 2019 and 2023 versions}

        \includegraphics[width=0.8\linewidth, keepaspectratio]{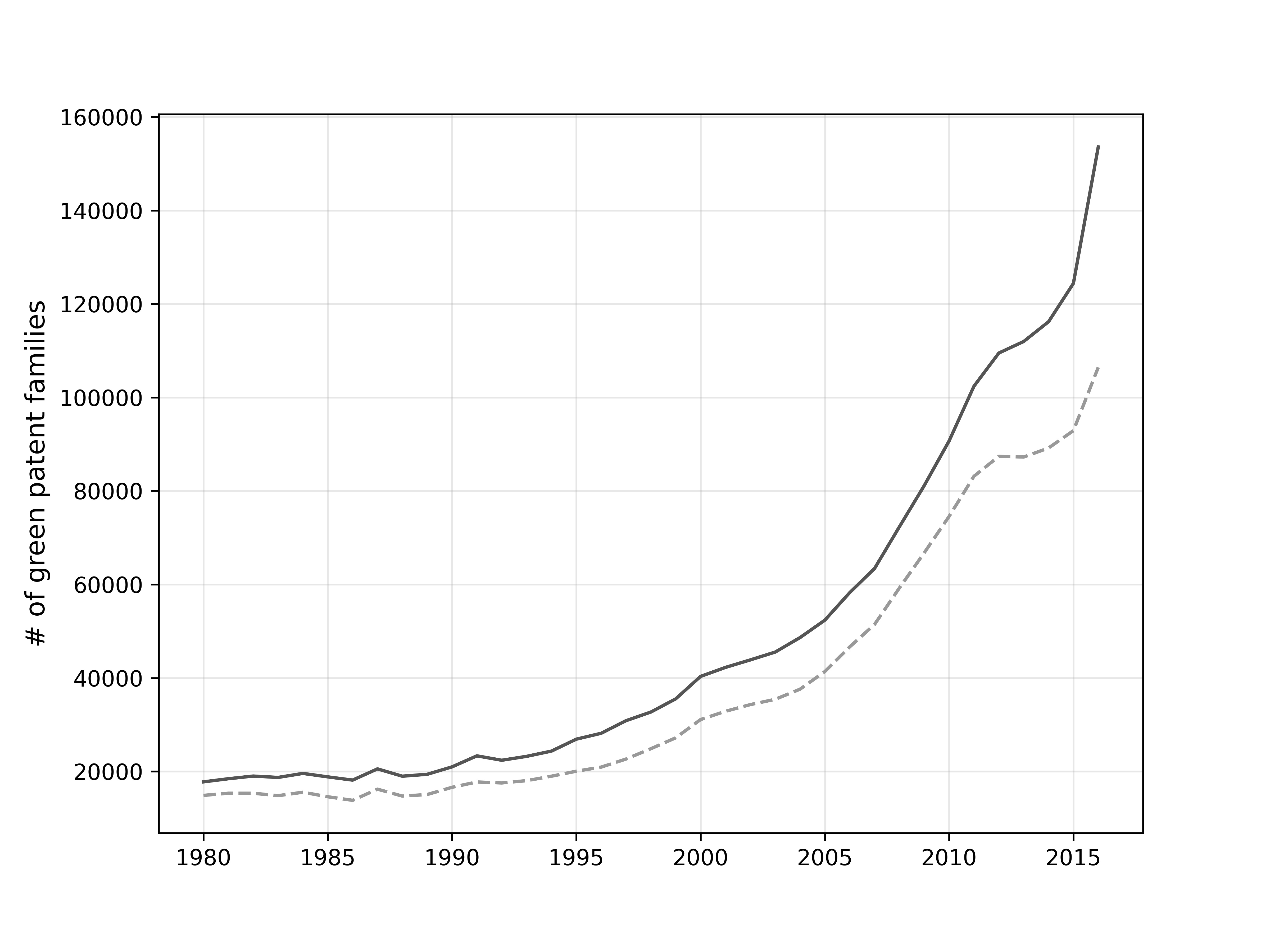}
        
\label{fig:baseline}
    \justifying \noindent \footnotesize
    
    Notes: The figure shows the number of green patent families (DOCDB) per year (earliest filing year). The gray dashed (solid black) line shows the family counts for the 2019 (2023) version.

\end{figure}

Figure ~\ref{fig:baseline} shows the number of green patent families calculated using PATSTAT 2019 (gray dashed line) and PATSTAT 2023 (black solid line). The two releases produce a similar broad trajectory: counts increase slowly during the 1980s and 1990s and accelerate from the early 2000s. The 2023 edition nevertheless records more green families throughout the observation period, and the difference between the two series becomes strong for recent priority years.

The divergence for recent years may partly reflect right truncation and delays in the publication and incorporation of patent records. These factors cannot, however, explain differences extending far back into the observation period. The gap may also result from reclassifications, expanded CPC and database coverage, and updates to patent family records. The baseline comparison therefore shows that database vintage affects both recent and historical estimates of green inventive activity. The next subsections decompose this difference into the mechanisms identified in the conceptual framework.
\footnote{As a descriptive check, we reconstructed the series after excluding patent families associated with China. The strong increase in recent priority years becomes substantially weaker, and both series peak around 2011–2012 before declining. This result illustrates the broader measurement issue examined in the paper: the observed temporal evolution of green patenting depends on the population of patents retained in the analysis. As shown more systematically in the filtering section, selection criteria that disproportionately exclude Chinese patent families can materially alter both the geographical composition and the temporal profile of the observed population. Nevertheless, the 2023 PATSTAT series remains consistently above the corresponding 2019 series even after China is excluded.}

 \FloatBarrier 
\subsection{The effects of reclassification, set expansion and filtering}


\FloatBarrier
\subsubsection{Reclassification}
We first examine reclassification. As defined in Section~\ref{sec:data_methods}, the analysis is restricted to patent families observed in both PATSTAT editions. A family is affected by reclassification when none of its members carries a Y02 symbol in the 2019 edition but at least one member carries a Y02 symbol in the 2023 edition. This comparison holds the observable family population constant and isolates the reclassification effects from coverage changes (i.e., set expansion effect).

\begin{figure}[h]
    \centering
        \caption{Patent families reclassified as green}

        \includegraphics[width=0.7\linewidth, keepaspectratio]{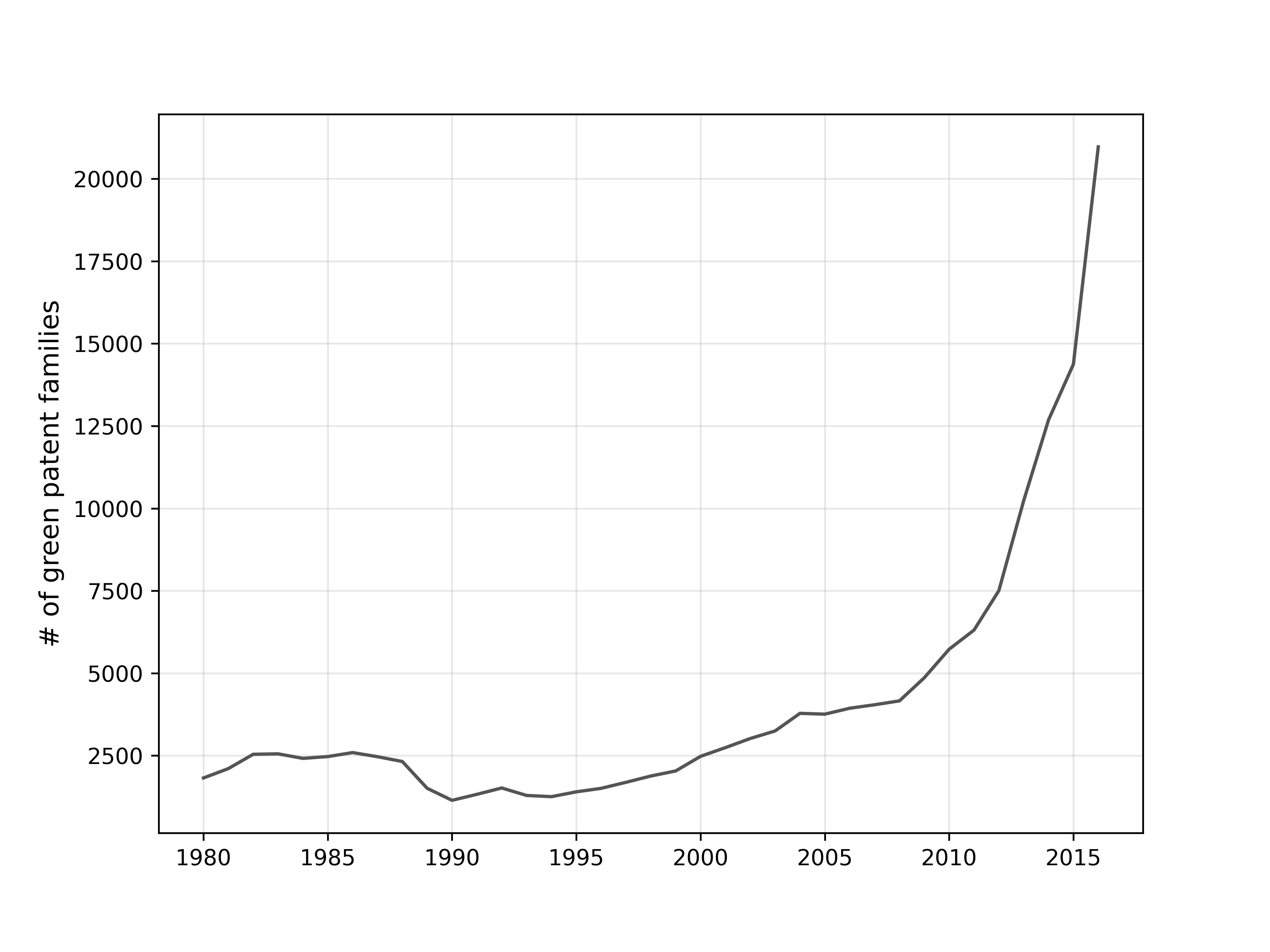}
        
\label{fig:recla_trends}
    \justifying \noindent \footnotesize
     Notes: The figure shows the number of green patent families (DOCDB) per year (earliest filing year) that experienced a reclassification. 
\end{figure}

Figure~\ref{fig:recla_trends} reports the number of families reclassified as green by earliest filing year. Reclassification remains comparatively limited for families originating in the 1980s and 1990s, increases from the early 2000s, and rises sharply after 2010. In 2016, more than 20,000 families were reclassified as green, representing over 16\% of the green families observed for that priority year. Reclassification is therefore particularly relevant for estimates of recent green inventive activity.

Across the full 1980--2016 period, 151,617 patent families are added to the observed green domain through reclassification. This corresponds to 9.2\% of green patents in the 2023 edition after excluding those attributed to set expansion. The corresponding reclassification rate across the broader CPC comparison reported in Appendix~\ref{app:reclass_class} is 2.2\%. Green reclassification is therefore about four times as frequent under this comparison. 

\begin{figure}
    \centering
        \caption{Reclassified green patent families by Y02 group}
        \includegraphics[width=0.9\linewidth, keepaspectratio]{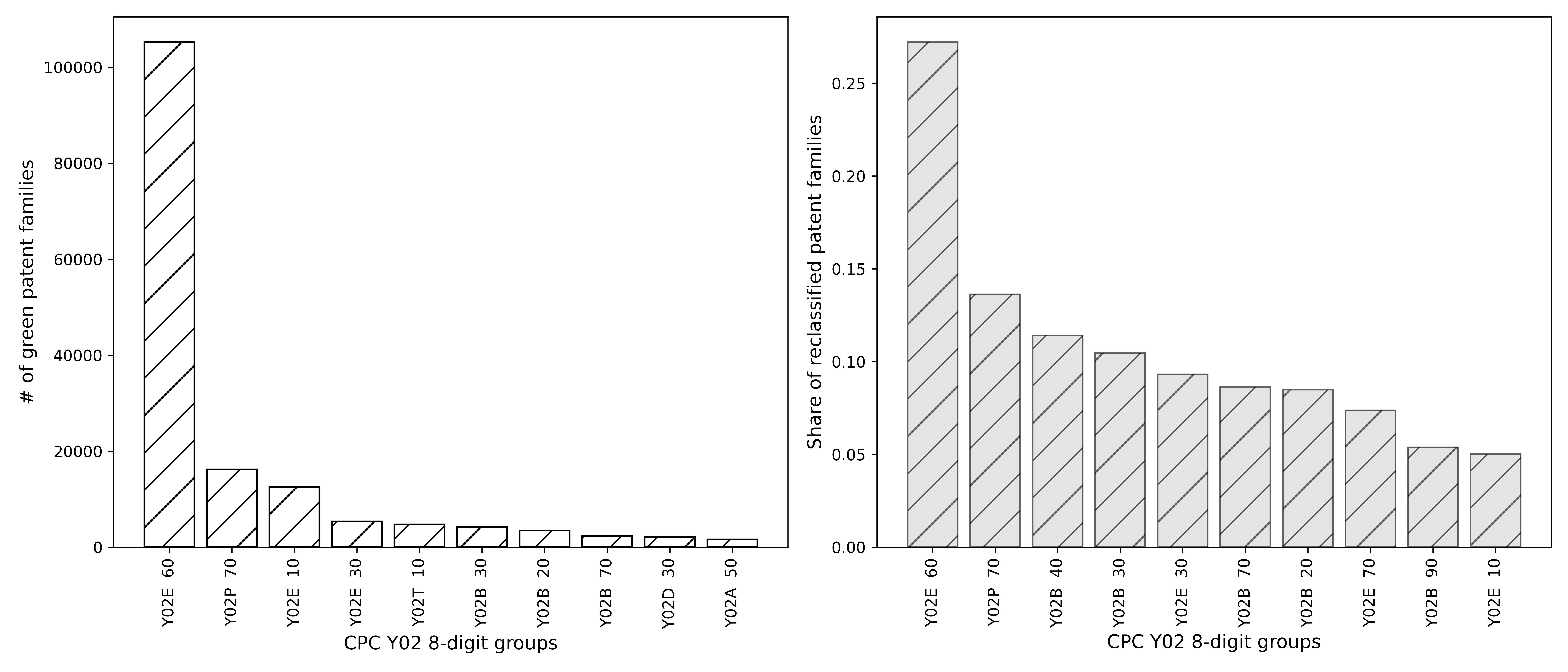}

\label{fig:recla_tech}
    \justifying \noindent \footnotesize
    Notes: The figure shows the distribution of reclassified patent families (DOCDB) by Y02 groups (8-digit CPC codes). In the left graph the bars show the total number of reclassified green patent families with an earliest filing year between 1980 and 2016. The right graph reports the share of reclassified green patent families over the total number of green patent families in the corresponding Y02 group. (Source: Own elaboration using data from PATSTAT 2019 and 2023).
    
\end{figure}

Next, we examine the qualitative nature of changes in the boundaries of green technology. 
Figure~\ref{fig:recla_tech} shows how reclassified families are distributed across Y02 groups, measured as absolute counts and shares of all green families assigned to the corresponding group. 

Energy-storage technologies classified under Y02E60 account for the largest number and the highest share of reclassified families, indicating that the observed boundary of green technology has changed extraordinarily strongly in this field. The group includes batteries, capacitors, fuel cells, and other technologies that support the storage and management of energy. All of them can be considered system-level technologies that support the integration of renewable energy into the broader system. 

Technologies for reducing greenhouse-gas emissions in the production of final industrial and consumer products (Y02P70) form the second-largest group, with approximately 18,000 reclassified families. Renewable energy generation technologies (Y02E10) and nuclear energy technologies (Y02E30) also account for substantial numbers of affected families. The absolute and normalized rankings differ because these technological groups vary considerably in size.

Reclassification is less intensive in most other Y02 groups. Road transport and building technologies contribute several thousand reclassified families, but their relative incidence is lower than in energy storage and production technologies. More generally, the results show that reclassification is concentrated in selected technological domains rather than distributed uniformly across the Y02 scheme.

\begin{figure}
    \centering

        \caption{Reclassified green patent families by patent office}

        \includegraphics[width=0.9\linewidth, keepaspectratio]{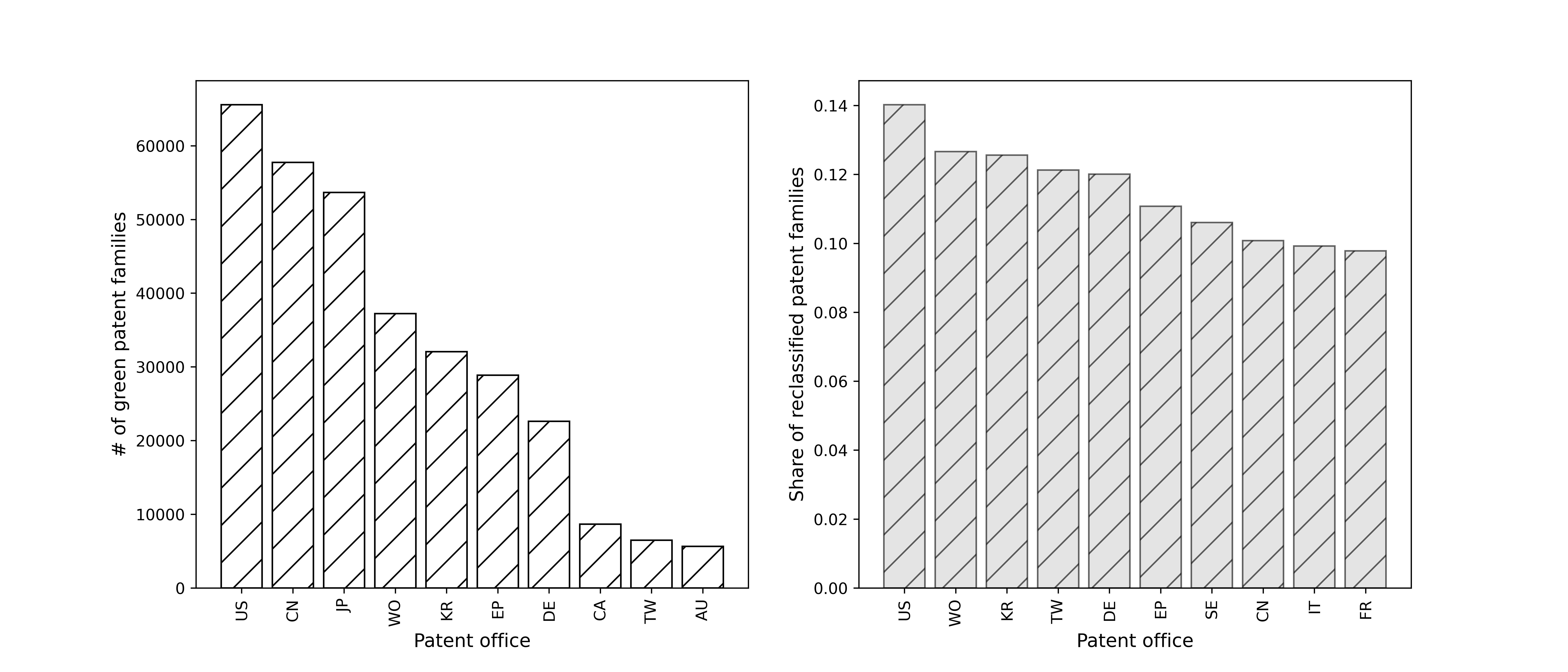}
        
\label{fig:recla_office}
    \justifying \noindent \footnotesize
    Notes: The figure shows the number of reclassified patent families (DOCDB) by patent office. A reclassified patent family is associated with an office if at least one member of the family had been filed at the office. `WO' denotes international applications filed under the Patent Cooperation Treaty. The data covers patent families filed between 1980 and 2016. 
\end{figure}

We now explore the institutional perspective and examine how reclassifications affect the distribution of observed green patenting across patent offices. 
Figure~\ref{fig:recla_office} examines the geographical distribution of reclassified families. A family is associated with an office when at least one of its members was filed there. The left panel reports the absolute number of reclassified families associated with each office, while the right panel reports their share of all green families associated with that office. Because a family may contain filings in several jurisdictions, office counts are not mutually exclusive.

In absolute terms (left panel), the largest numbers of reclassified families are associated with the USPTO, the Chinese (CN) and Japanese (JP) offices. International Patent Cooperation Treaty (PCT) (WO) and Korean (KR) filings follow, while the EPO and German (DE) patent office also account for substantial numbers. These rankings partly reflect the overall volume of patenting at the respective offices. 

The normalized results (right panel) provide a different perspective. Reclassified families account for the largest share of green families at the USPTO, followed by patents filed through the PCT route  and Korean office. The Chinese office ranks high in absolute terms but lower in relative terms because its total population of green patent families is substantially larger. Taiwan, Germany, and Sweden also display relatively high reclassification shares. Reclassification therefore appears to exert more limited effects on the geographical distribution of green inventive activity, with slightly different effects depending on whether measuring green patenting in absolute or relative terms.

\FloatBarrier
\subsubsection{Set expansion}

\begin{figure}
    \centering
        \caption{Green patent families attributed to set expansion}

        \includegraphics[width=0.7\linewidth, keepaspectratio]{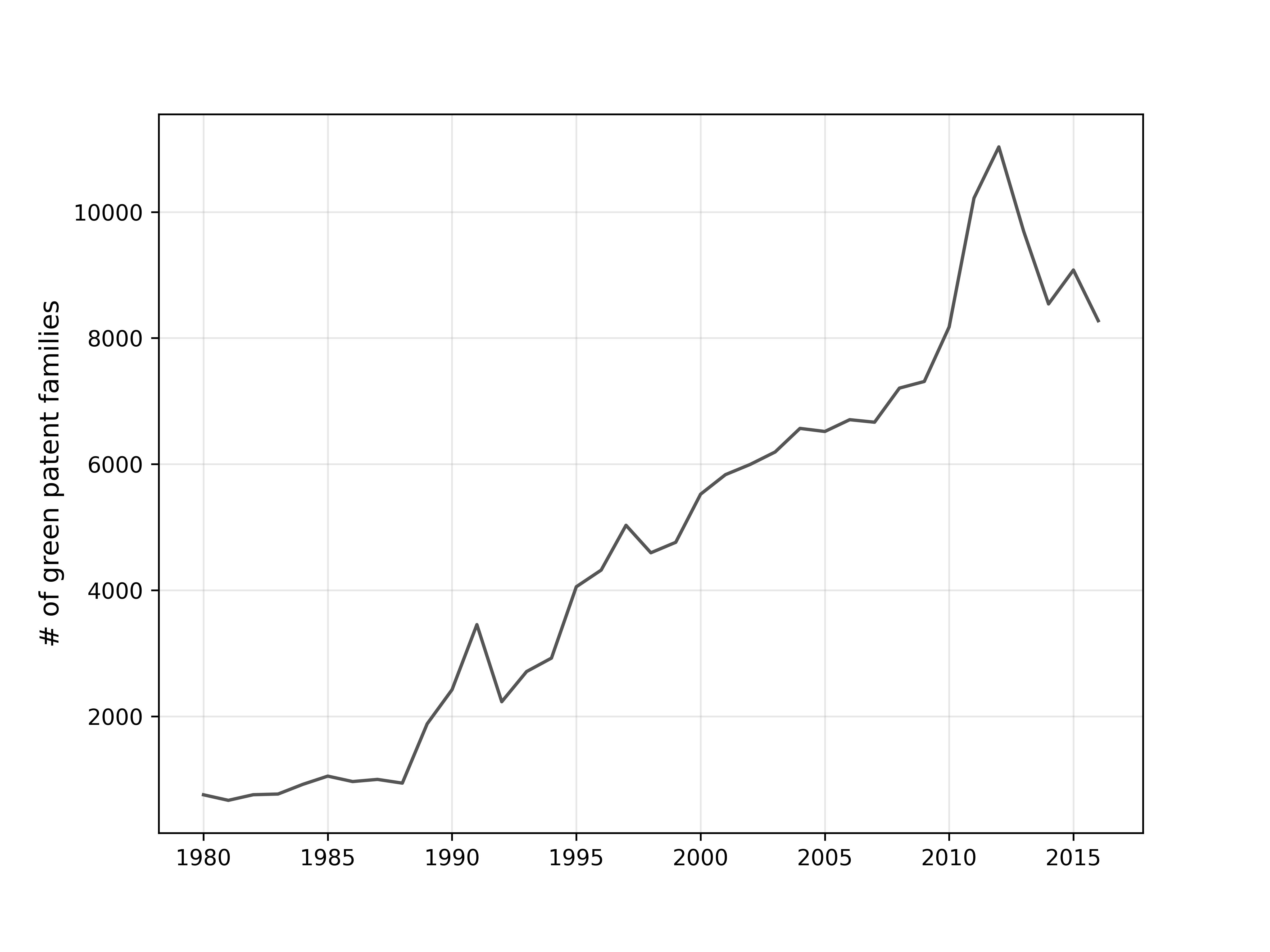}        
\label{fig:set_trend}

    \justifying \noindent \footnotesize
    Notes: The figure shows the number of green patent families (DOCDB) being added to the 2023 version by earliest filing year and classified as green according to the 2023 CPC codes. (Source: Own elaboration using data from PATSTAT 2019 and 2023).

\end{figure}

Figure~\ref{fig:set_trend} reports the number of green patent families added through set expansion by earliest filing year. Counts remain comparatively low for families originating in the 1980s, increase through the 1990s and 2000s, and rise sharply around 2010 before reaching a peak around 2013. They decline for the most recent priority years, yet the underlying records, family structures, and retrospective CPC coverage may still be incomplete. 
Across the full observation period, set expansion adds 175,732 green patent families to the green sample observed in the 2023 release, corresponding to 10.6\% of the 2023 sample.

\begin{figure}
    \centering
        \caption{Green patent families attributed to set expansion by technological domain}

        \includegraphics[width=0.9\linewidth, keepaspectratio]{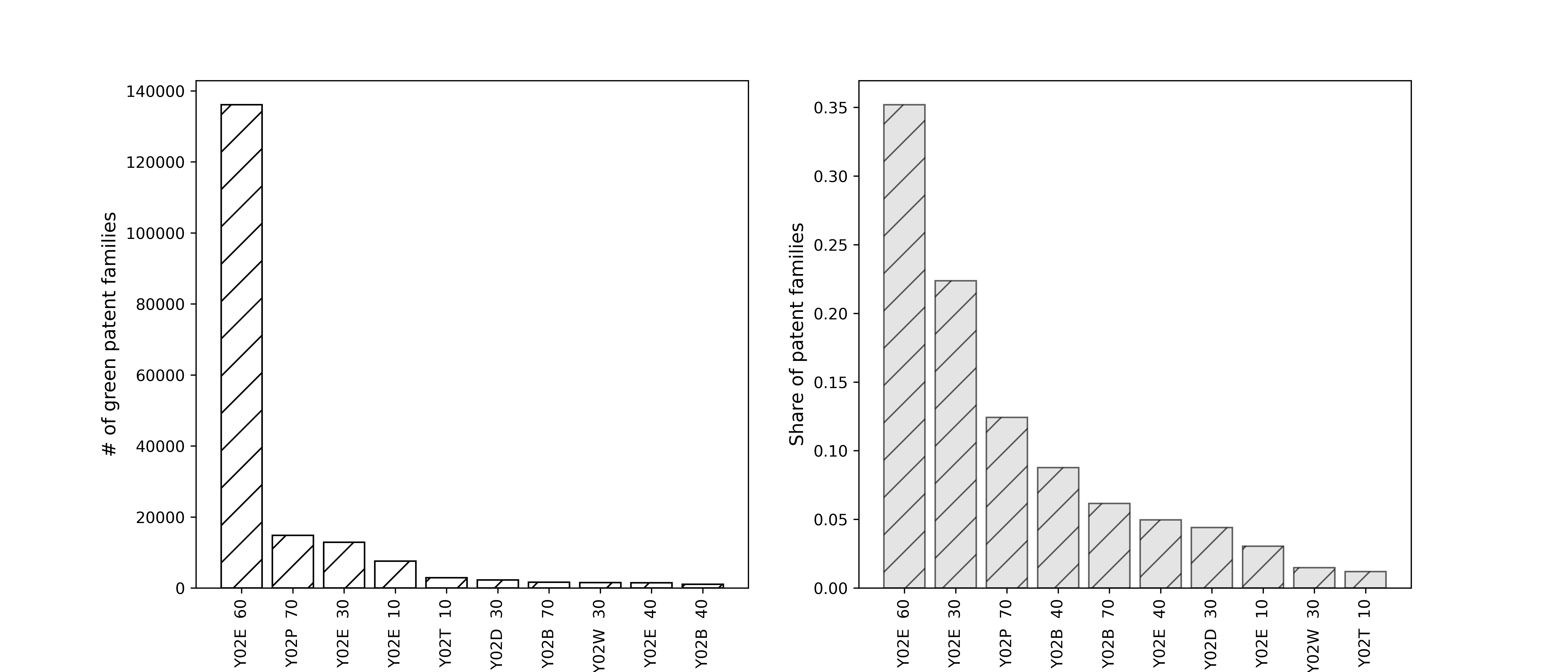}

\label{fig:set_office_tech}

    \justifying \noindent \footnotesize
    Notes: The figure shows the number of green patent families (DOCDB) being added through the set expansion by Y02 classes. The data covers patent families filed between 1980 and 2016 and patent families are classified as green using the 2023 version of CPC codes. 
    
\end{figure}

We first look at the potential uneven impact across green technological domains. Figure~\ref{fig:set_office_tech} shows that set expansion follows a pattern similar to the reclassification effect, that is, it is highly concentrated across green technological domains rather than being uniformly distributed throughout the Y02 scheme. In absolute terms, energy storage technologies (Y02E60) dominate, accounting for a disproportionately large share of patent families. Renewable energy technologies (Y02E10) and technologies for reducing emissions in the production of final goods (Y02P70) follow, although at substantially lower levels. The normalized results reinforce, but also qualify, this pattern. Around 35\% of all patent families under Y02E60 were added to the green domain through the set expansion effect. Nuclear energy (Y02E30) ranks only fourth in absolute numbers but second in relative terms, with more than one fifth of the corresponding green patent population affected. The figure suggests that set expansion disproportionately reshapes the visibility of energy storage, nuclear energy, low-carbon production, and selected building and transport technologies.

\begin{figure}
    \centering
        \caption{Green patent families attributed to set expansion by patent office}

        \includegraphics[width=0.9\linewidth, keepaspectratio]{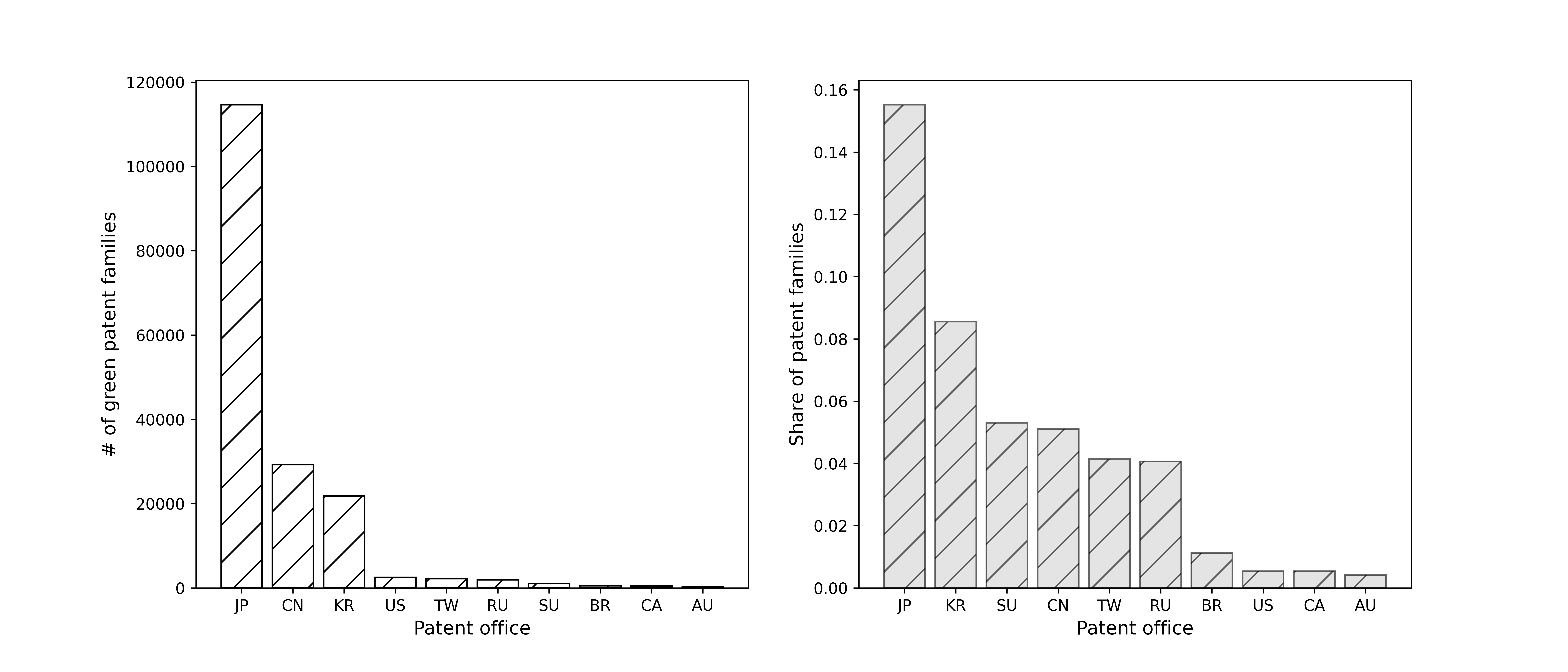}

\label{fig:set_office}

    \justifying \noindent \footnotesize
    Notes: The figure shows the number of green patent families (DOCDB) being added through the set expansion by patent office. A green patent family is associated with an office if at least one member of the family had been filed in the jurisdiction. The data covers patent families filed between 1980 and 2016 and patent families are classified as green using the 2023 version of CPC codes. 
    
\end{figure}

Second, Figure~\ref{fig:set_office} shows that set expansion is also geographically concentrated. The largest numbers of affected families are associated with the Japanese, Chinese, and Korean patent offices. Because a family is associated with every office in which at least one family member was filed, office counts are not mutually exclusive. 

CPC coverage expanded unevenly across patent offices after the system was introduced in 2013. Some offices initially classified only parts of their current documentation and the retrospective CPC coverage of existing patents developed gradually \citep{DEGROOTE2018S78}. Appendix~\ref{app:cpc_coverage} documents these differences for the major Asian offices.  
Coverage was comparatively comprehensive for EPO and USPTO documentation, whereas CPC coverage of Japanese, Korean, and Chinese records remained substantially less complete during much of the 2010s, but, in some cases, increased sharply over the years. \citep{DEGROOTE2018S78}. 
These patterns provide an institutional explanation for why inventions associated with particular jurisdictions may become observable only in later PATSTAT editions.

The right panel normalizes these counts by the total number of green patent families associated with each office. The concentration of set expansion in the major Asian offices indicates that their contributions to the green transition were underrepresented in earlier releases of PATSTAT. Consequently, changes in measured green technological performance across editions may partly reflect expanded institutional and database coverage rather than contemporaneous changes in inventive activity.

These results highlight that the set expansion reshapes both the technological and geographical distribution of green inventions.

\FloatBarrier
\subsubsection{Filtering}
\label{subsec:results_quality}

We next examine how user-imposed filtering criteria, aimed at identifying high-value inventions, change the observed green technological landscape as captured by patents. 
These criteria capture different dimensions of technological influence, international protection, and institutional coverage. 
The family size filter retains patent families containing applications in at least two jurisdictions. The citation filter selects families receiving at least one forward citation within the five-year citation window defined in Section~\ref{sec:data_methods}. The EPO and USPTO filters retain families with at least one application filed at the respective office, while the triadic filter gathers families containing applications at the EPO, USPTO, and Japan Patent Office \citep{criscuolo2006home, squicciarini2013measuring, higham2021patent}. 

\begin{figure}[htbp]

    \caption{Green patent families under alternative filters}

    \centering
        \begin{subfigure}{0.45\textwidth}
        \includegraphics[width=\linewidth, height=0.45\textheight, keepaspectratio]{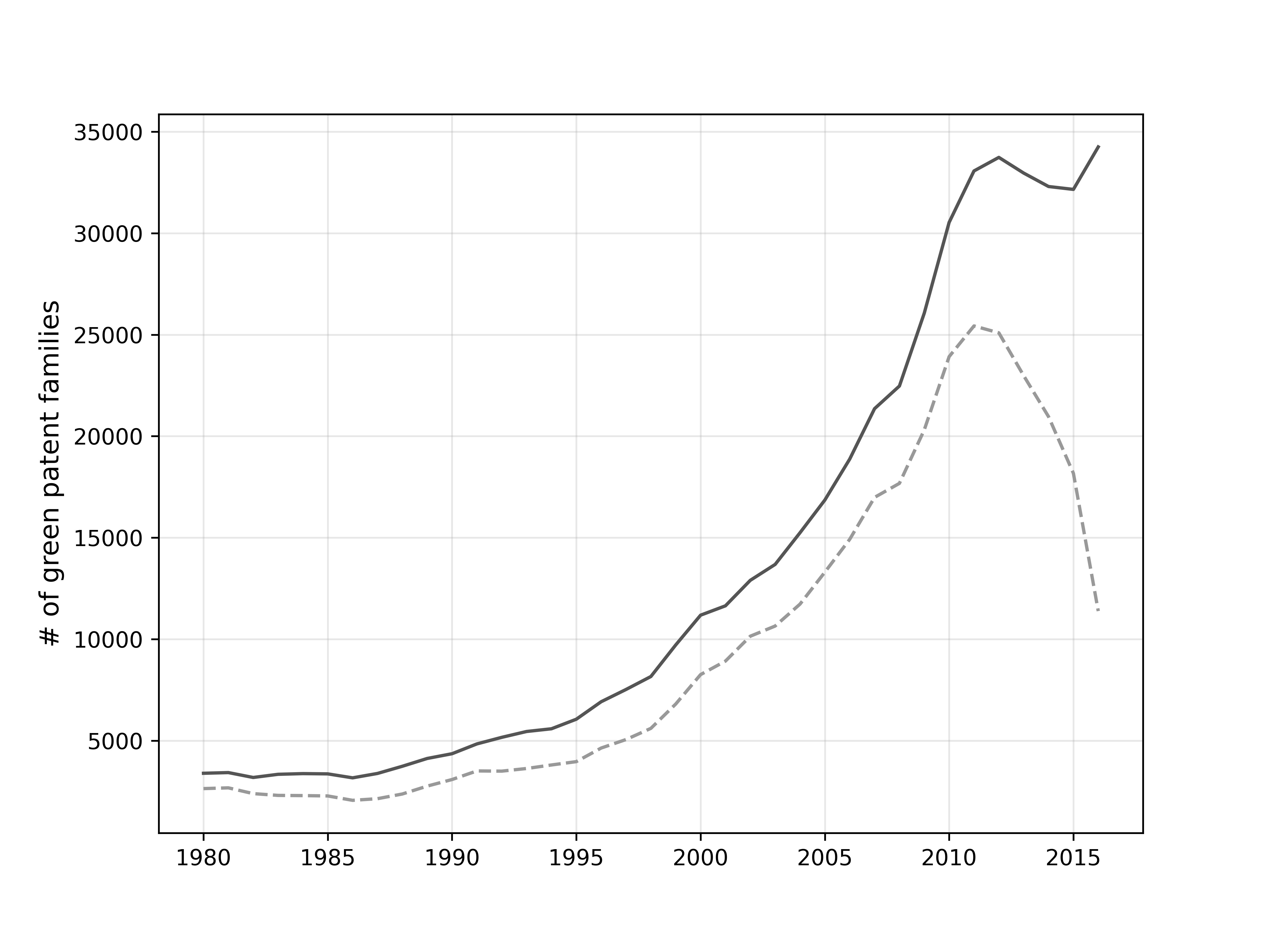}
        \caption{Family size $>$1}
        \label{subfig:trend_famsize}
    \end{subfigure}
    \hfill
    \begin{subfigure}{0.45\textwidth}
        \includegraphics[width=\linewidth, height=0.45\textheight, keepaspectratio]{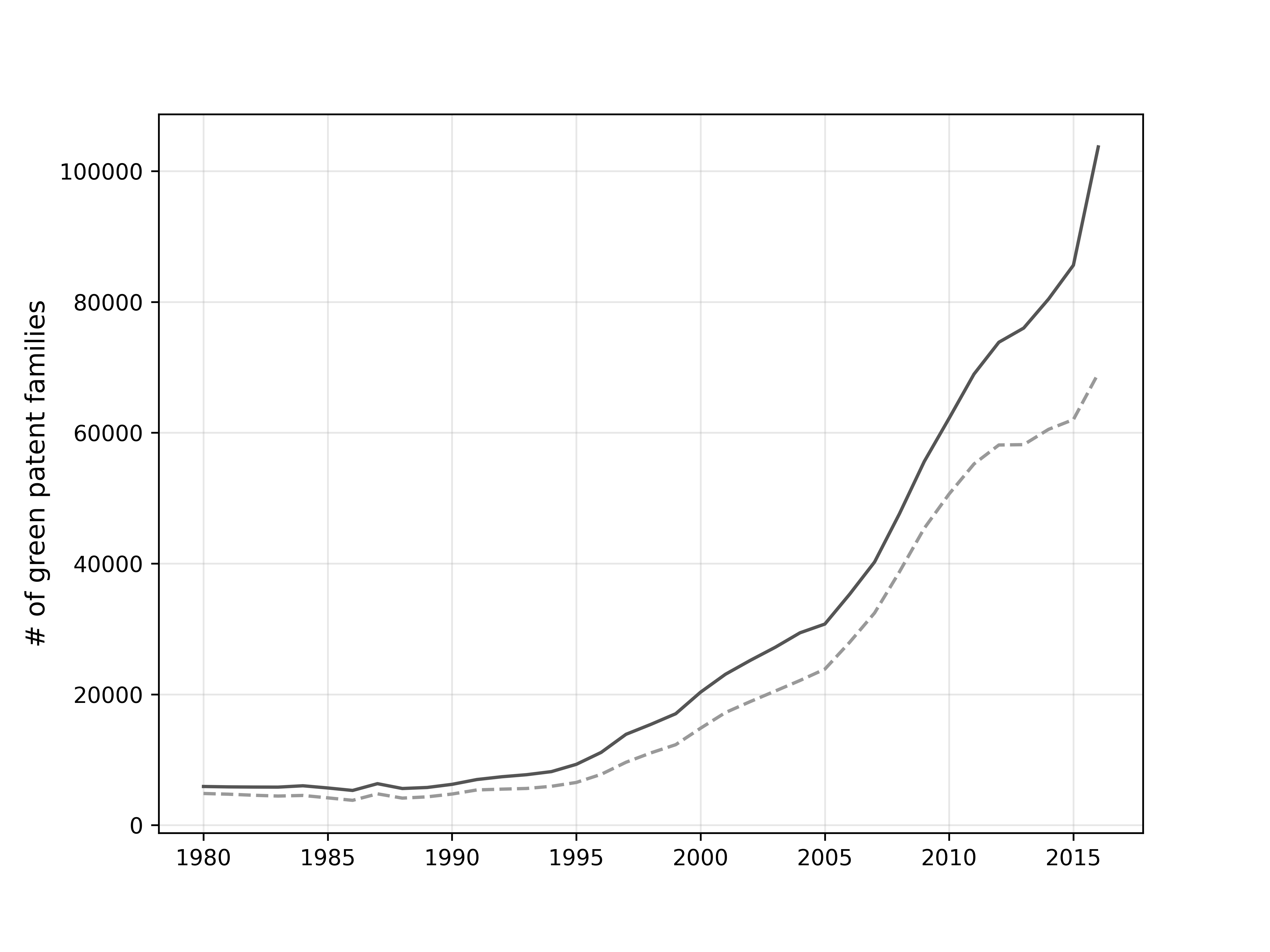}
        \caption{Forward citations $>$0}
        \label{subfig:trend_cit1}
    \end{subfigure}
    
    
    \begin{subfigure}{0.45\textwidth}
        \includegraphics[width=\linewidth, height=0.45\textheight, keepaspectratio]{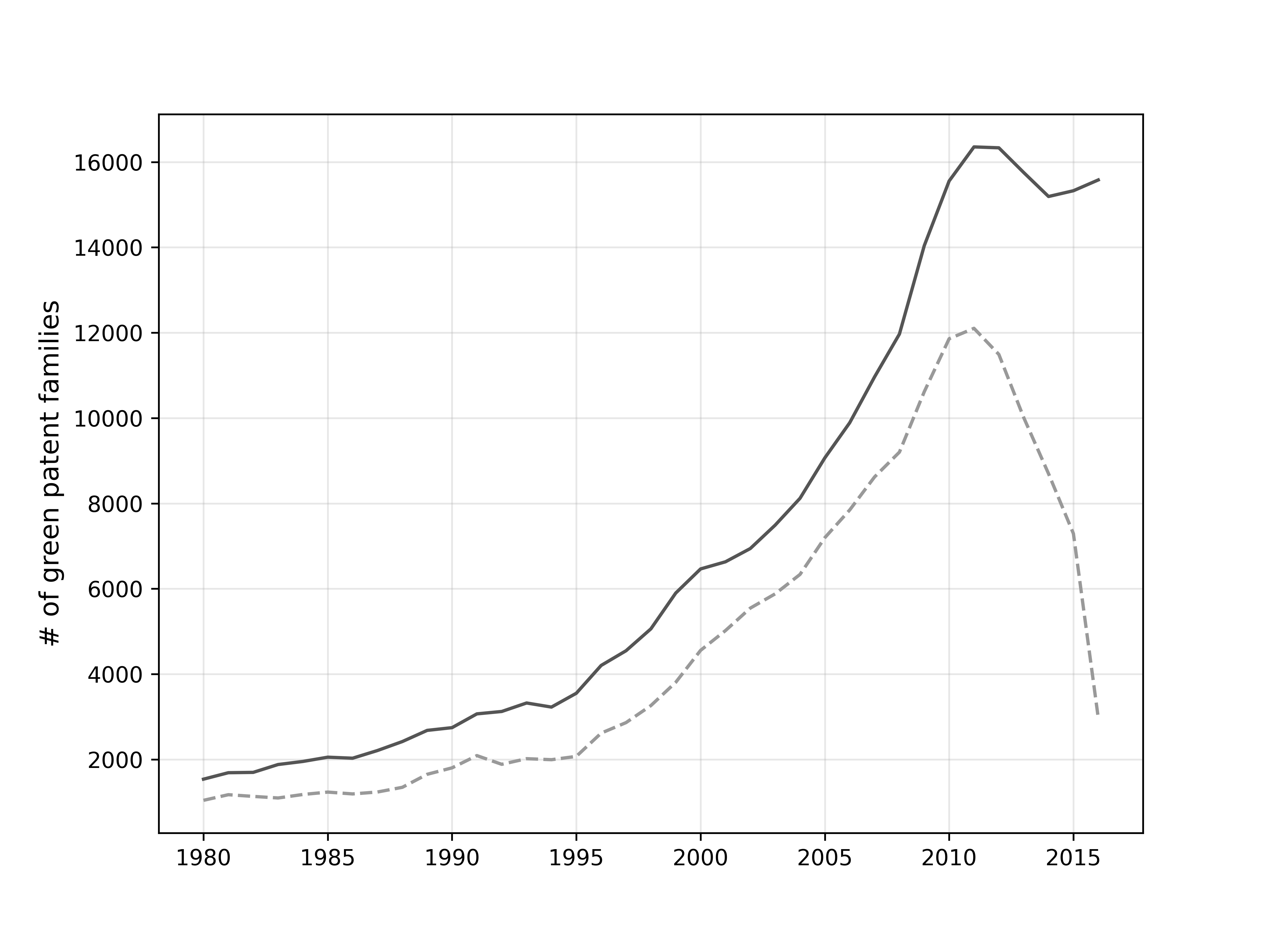}
        \caption{EPO family component}
        \label{subfig:trend_ep}
    \end{subfigure}
    \hfill
    \begin{subfigure}{0.45\textwidth}
        \includegraphics[width=\linewidth, height=0.45\textheight, keepaspectratio]{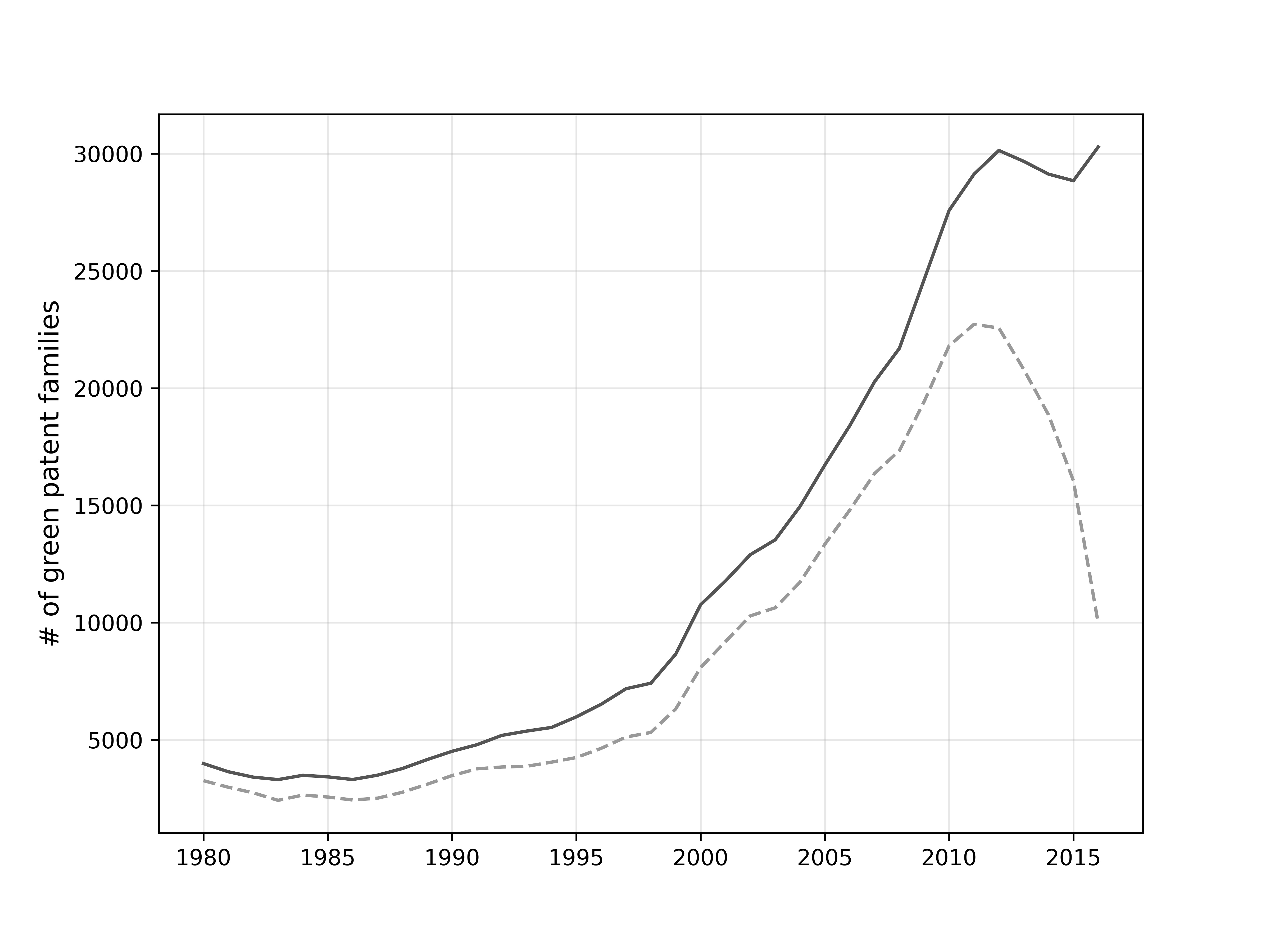}
        \caption{USPTO family component}
        \label{subfig:trend_us}
    \end{subfigure}

    
    \begin{subfigure}{0.45\textwidth}
        \includegraphics[width=\linewidth, height=0.45\textheight, keepaspectratio]{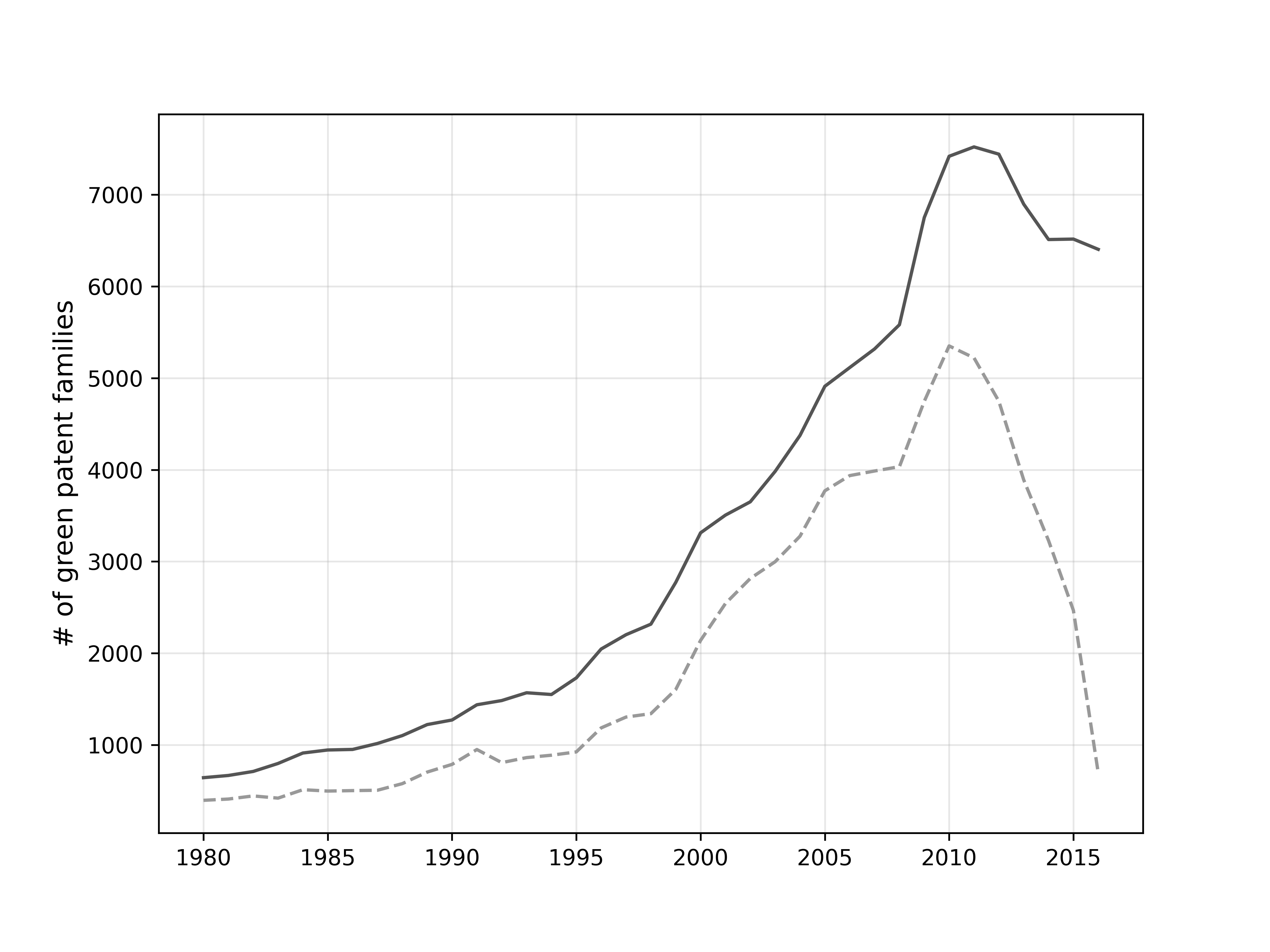}
        \caption{Triadic patent families}
        \label{subfig:trend_triadic}
    \end{subfigure}
    \hfill

    \label{fig:trend_quality}
    \justifying \noindent \footnotesize
    Notes: The figure reports the trends in green patent families (DOCDB) over the period 1980-2016 (earliest filing year). Each graph applies the corresponding patent quality filter. (Source: Own elaboration using data from PATSTAT 2019 and 2023).
    
\end{figure}

Figure~\ref{fig:trend_quality} reports annual green patent family counts under each filter. The gray dashed lines use the Spring 2019 PATSTAT edition, and the solid black lines use the Spring 2023 edition. Comparing the panels with the unfiltered baseline in Figure~\ref{fig:baseline} highlights how both the level and temporal patterns of green inventive activity depend on the filtering criterion. 
Because the indicators are calculated separately in each release, differences between the two lines can also reflect changes of the underlying information. For example, patent families may acquire additional members or additional citations between two database vintages. Filtering is therefore itself sensitive to database release.

The family size, EPO, USPTO, and triadic filters (Figures ~\ref{subfig:trend_famsize}, \ref{subfig:trend_ep}-\ref{subfig:trend_triadic}) show similar trajectories in the 2019 edition. Counts remain comparatively low until the late 1990s, increase rapidly during the 2000s, reach a peak around 2012, and subsequently decline. The magnitude of the series differs substantially across filters, with the triadic criterion retaining the smallest population.

The 2023 edition modifies the post-2012 pattern. Under the family size, EPO, and USPTO filters, the decline weakens and we observe an increase again around 2015. No comparable recovery is visible for triadic families. Thus, the persistence of the post-2012 slowdown depends jointly on the database vintage and the filtering threshold imposed on the analytical sample.

The citation filter (Figure~\ref{subfig:trend_cit1}) produces a different dynamic. Green patent counts increase throughout most of the period and accelerate during the early 2000s. The 2023 edition yields consistently higher counts than the 2019 release because it contains a larger and more mature citation network, allowing additional patents to cross the threshold of at least one forward citation. 

This comparison requires caution. The 2019 release cannot observe the same complete five-year citation window as the 2023 edition, especially for patents with recent priority date. Part of the divergence between the two series is therefore mechanically attributable to truncation rather than vintage effects.

The filters produce different estimates of the scale and trends of green inventive activity. Family-, office-, and triadic-based criteria substantially reduce the observed population and yield different conclusions about the persistence of the post-2012 slowdown. Citation-based filtering retains a broader geographical distribution but may be sensitive to the time for citations to accumulate. Hence, filter choice can be a substantive element of the research design and is not a neutral pre-processing decision.

\begin{figure}
    \centering
        \caption{Green patent families by technological domain and filter}

        \includegraphics[width=\linewidth, keepaspectratio]{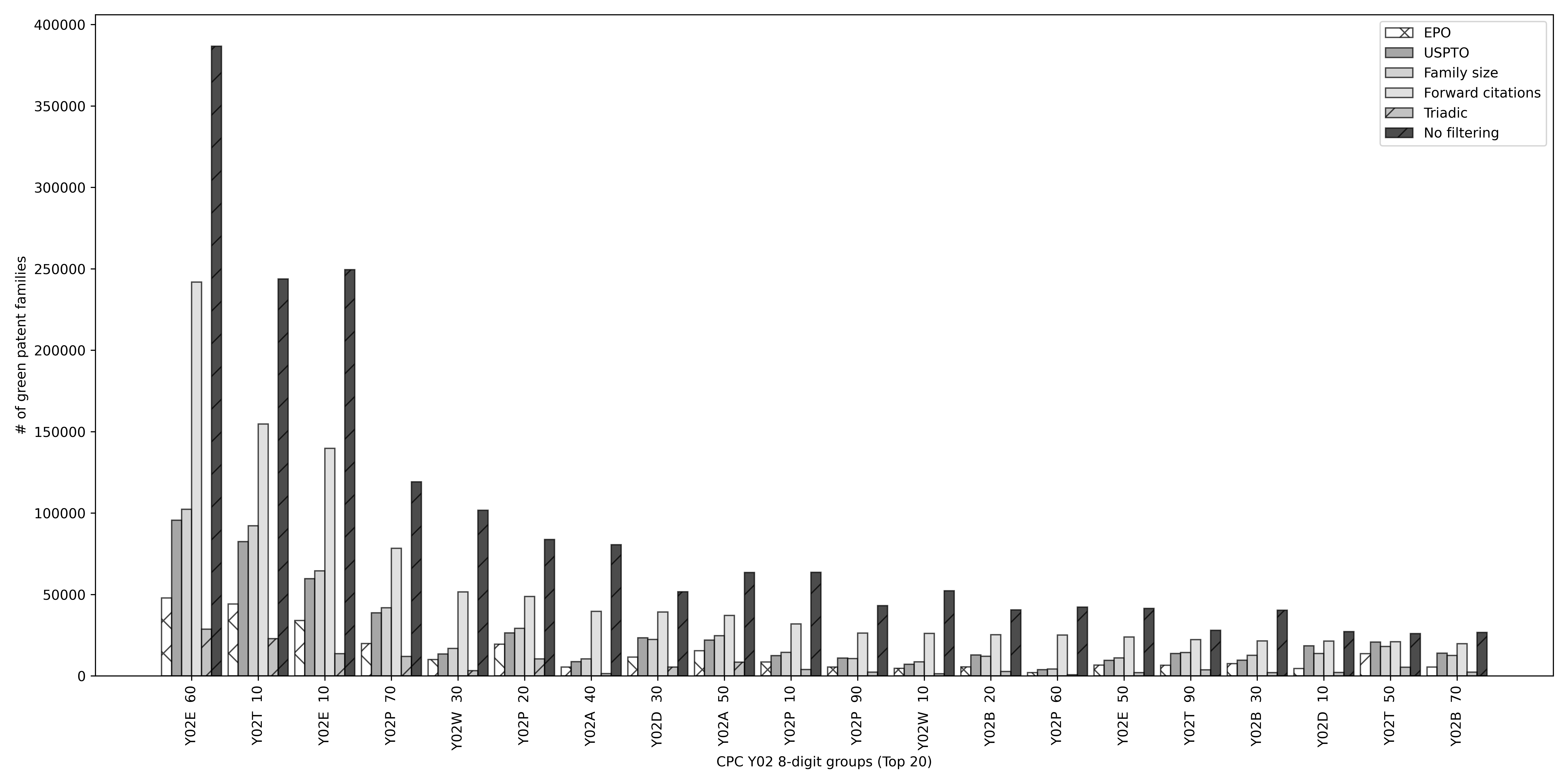}
        
    \label{fig:filtering_tech_all}
    \justifying \noindent \footnotesize
    \justifying\noindent\footnotesize
    \textit{Notes}: The figure reports green DOCDB patent family counts by Y02 class under the alternative filters. The sample covers earliest filing years 1980--2016. 
    
\end{figure}

We now examine how filtering affects the representation of technological domains. Figure~\ref{fig:filtering_tech_all} shows that the main effect is quantitative: all filters substantially reduce the number of observed patent families, although the magnitude of this reduction varies across fields and is particularly strong under office-based and triadic criteria. The overall technological hierarchy, however, remains broadly stable. The largest domains in the unfiltered sample generally remain among the most prominent after filtering, while changes are mainly confined to limited reversals between similarly sized fields. For example, Y02T10 overtakes Y02E10 under the citation and family-size filters, and Y02B30 moves above Y02P60 under some criteria. Filtering therefore affects the measured scale of technological activity more strongly than the overall ranking of green technological domains. Its use may still change the relative weight assigned to specific fields, but it does not substantially alter the broad structure of the observed green technology landscape.

\begin{figure}
    \centering
        \caption{Green patent families by patent office and filter}

        \includegraphics[width=\linewidth, keepaspectratio]{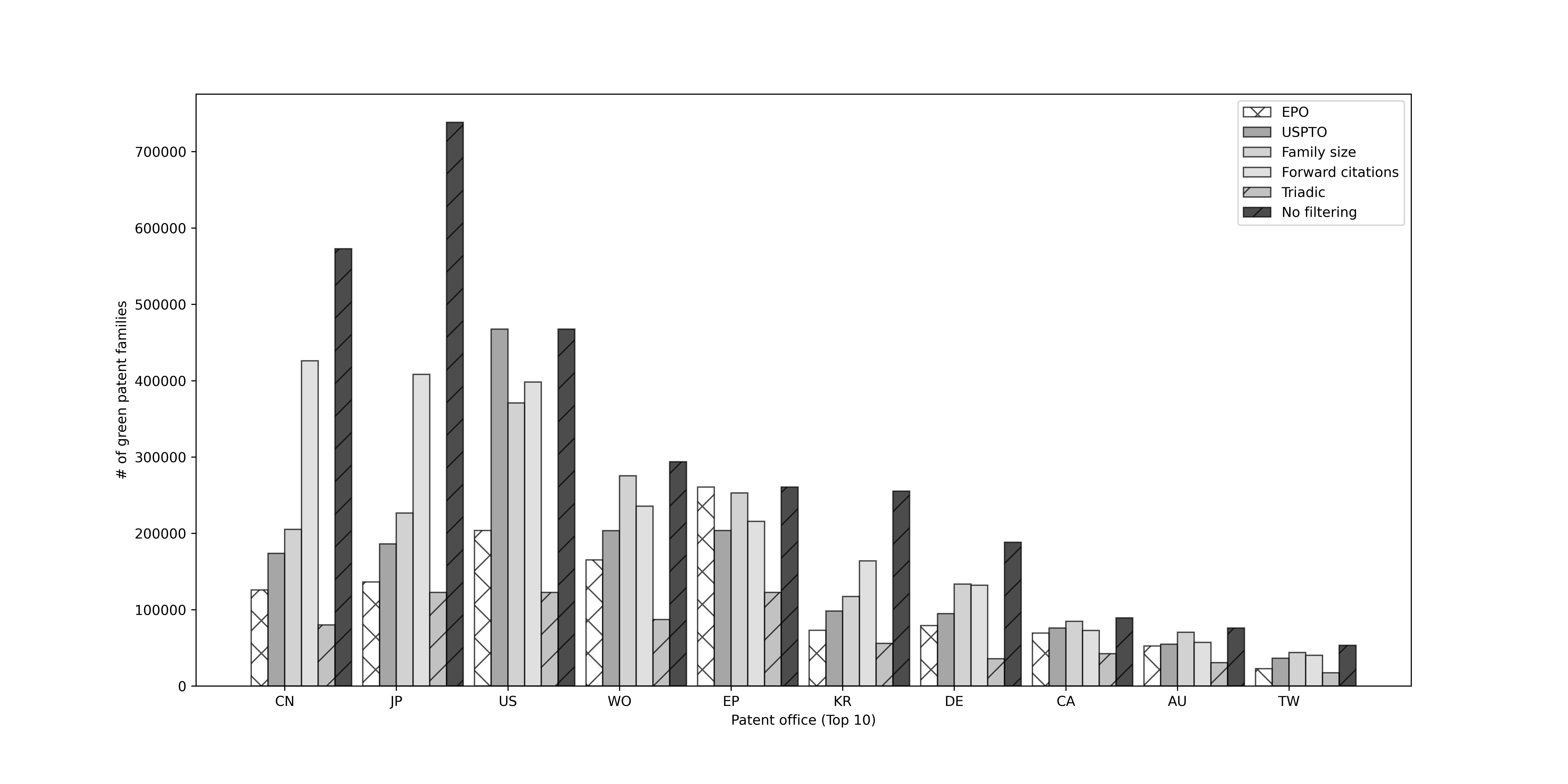}
        
    \label{fig:filtering_office_all}
    \justifying \noindent \footnotesize
    \justifying\noindent\footnotesize
    \textit{Notes}: The figure reports green DOCDB patent family counts by patent office under the alternative filters. A family is associated with an office if at least one family member was filed there; families may therefore be counted for more than one office. The sample covers earliest filing years 1980--2016. (Source: Own elaboration using data from PATSTAT 2019 and 2023).

\end{figure}

Figure~\ref{fig:filtering_office_all} compares the distribution of green families across patent offices under alternative filters. The distribution produced by the citation filter most closely resembles that of the unfiltered sample, with a relatively strong prominence of the Chinese, Japanese, and Korean offices. Family size-, office-, and triadic-based filters alter the distribution more substantially because they select directly or indirectly by certain jurisdictions.\footnote{Restricting the comparison to families with earliest filing years between 2010 and 2016 makes the geographical differences across filters more pronounced. Results are available upon request.}

The EPO and USPTO populations are comparatively less sensitive to most filters, partly because office-based and international-family criteria are closely aligned with filing through these established patent systems. The triadic criterion remains highly restrictive for all jurisdictions because it requires protection through three major patent offices.

Overall, filtering slightly alters the technological composition of green innovation while substantially reshaping its geographical representation, particularly by reducing the visibility of patent families associated with Asian offices and favouring inventions connected to established international filing systems.


\FloatBarrier
\subsection{Interactions between the mechanisms}

Table~\ref{tab:combination} examines how filtering interacts with reclassification and set expansion. Each row applies the indicated filter separately to the green samples observed in the 2019 and 2023 PATSTAT editions. The final two columns report the sizes of reclassified and set expansion samples after filtering. 

The interaction differs strongly across reclassification and set expansion effects. Set expansion samples account for 9.7\% of the unfiltered 2023 population and 8.3\% of the citation-filtered population. By contrast, less than 1\% of the family-size-, triadic-, EPO-, and USPTO-filtered samples consists of set expansion families. These filters therefore remove most of the inventions that become observable through expanded database and CPC coverage.

Reclassification remains quantitatively important under every filter. Reclassified families account for between 8.4\% and 16.4\% of the corresponding 2023 samples, with especially large shares under the USPTO, triadic, and family-size criteria. Reclassified families are therefore considerably more likely than set expansion families to remain visible after international-family and office-based restrictions are imposed.

The filters under which the post-2012 decline remains most pronounced are also those that retain very few set expansion families. This pattern suggests that the persistence of the measured slowdown partly reflects the exclusion of inventions associated with expanded institutional coverage. However, this does not mean that the set expansion is the sole cause of the observed differences between the trends.

\begin{table}[h!]
\centering
\caption{Reclassification and set expansion samples under alternative filters}


\begin{tabular}{lrrrr}
\toprule
\textbf{Filter} & \textbf{2019} & \textbf{2023} & \textbf{Reclassification} & \textbf{Set expansion} \\
\midrule
No filtering & 1,412,363 & 1,814,580 & 151,617 (8.4\%) & 175,732 (9.7\%) \\
Citations & 794,349 & 1,046,702 & 101,713 (9.7\%) & 87,104 (8.3\%) \\
Family size & 348,302 & 497,510 & 65,263 (13.1\%) & 3,225 (0.65\%) \\
Triadic & 75,469 & 122,563 & 16,171 (13.2\%) & 41 (0.03\%) \\
EPO & 171,979 & 260,596 & 28,863 (11.1\%) & 131 (0.05\%) \\
USPTO & 296,358 & 400,143 & 65,551 (16.4\%) & 2,519 (0.63\%) \\
\bottomrule
\end{tabular}

\label{tab:combination}

\justifying\noindent\footnotesize
\textit{Notes}: The first two columns report green DOCDB patent family counts after applying each filter separately in the 2019 and 2023 PATSTAT editions. The final two columns report the numbers of families attributed to reclassification and set expansion that satisfy the
corresponding filter. Percentages in parentheses are relative to the filtered 2023 population. 
The mechanism-specific columns are not intended to sum to the difference between editions.

\end{table}


    \subsection{The nature of excluded inventions}
\label{subsec:patent_characteristics_results}

The previous sections showed that the reclassification, set expansion, and filtering effects substantially alter the number, technological composition, and geographical distribution of green patent families. However, these effects may not be randomly distributed across inventions. If excluded patents systematically differ from those that remain observable, changes in the information infrastructure may shape not only the measured volume of green innovation but also how its characteristics are interpreted within the green technological domain.

\begin{table}[htbp]
\centering
\caption{Patent characteristics across mechanism-specific samples}
\label{tab:excluded_patent_characteristics}
\footnotesize
\setlength{\tabcolsep}{4.5pt}
\begin{tabular}{lccccc}
\toprule
& \multicolumn{2}{c}{\textbf{Complexity}}
& \multicolumn{1}{c}{\textbf{Novelty}}
& \multicolumn{2}{c}{\textbf{Impact}} \\
\cmidrule(lr){2-3}
\cmidrule(lr){4-4}
\cmidrule(lr){5-6}
& \makecell{Scope\\(4-digit)} 
& Originality 
& \makecell{Novelty\\ in \\ recombination \\ (4-digit)}
& \makecell{Forward \\ Citations\\(5 years)} 
& Generality \\
\midrule

\multicolumn{6}{l}{\textit{Panel A. Reclassification}} \\

Affected by reclassification
& $-0.0629^{***}$
& $-0.0283^{***}$
& $ 0.0028^{***}$
& $ 0.1355^{***}$
& $-0.0276^{***}$ \\
& $(0.0012)$
& $(0.0005)$
& $(0.0002)$
& $(0.0096)$
& $(0.0006)$ \\

Observations
& 3,314,059
& 2,686,858
& 3,314,059
& 3,314,059
& 2,362,918 \\

Pseudo $R^2$ / $R^2$
& 0.0289
& 0.1408
& 0.0194
& 0.0983
& 0.0977 \\[8pt]

\multicolumn{6}{l}{\textit{Panel B. Set expansion}} \\

Affected by set expansion
& $-0.1233^{***}$
& $-0.0411^{***}$
& $ 0.0035^{***}$
& $ 0.1484^{***}$
& $-0.0545^{***}$ \\
& $(0.0017)$
& $(0.0011)$
& $(0.0002)$
& $(0.0101)$
& $(0.0011)$ \\

Observations
& 3,119,837
& 2,410,122
& 3,119,837
& 3,119,837
& 2,151,692 \\

Pseudo $R^2$ / $R^2$
& 0.0288
& 0.1330
& 0.0203
& 0.1093
& 0.0946 \\[8pt]

\multicolumn{6}{l}{\textit{Panel C. EP-office filtering}} \\

Excluded by EP-office filter
& $-0.0204^{***}$
& $-0.0139^{***}$
& $-0.0008^{***}$
& $ 0.0524^{***}$
& $ 0.0045^{***}$ \\
& $(0.0009)$
& $(0.0004)$
& $(0.0001)$
& $(0.0080)$
& $(0.0005)$ \\

Observations
& 2,924,122
& 2,327,758
& 2,924,122
& 2,924,122
& 2,053,054 \\

Pseudo $R^2$ / $R^2$
& 0.0284
& 0.1257
& 0.0202
& 0.1099
& 0.0870 \\[8pt]

\multicolumn{6}{l}{\textit{Panel D. US-office filtering}} \\

Excluded by US-office filter
& $-0.5176^{***}$
& $-0.3435^{***}$
& $-0.0064^{***}$
& $-1.2045^{**}$
& $-0.0979$ \\
& $(0.0583)$
& $(0.0727)$
& $(0.0008)$
& $(0.5364)$
& $(0.0709)$ \\

Observations
& 2,924,122
& 2,327,758
& 2,924,122
& 2,924,122
& 2,053,054 \\

Pseudo $R^2$ / $R^2$
& 0.0284
& 0.1257
& 0.0202
& 0.1099
& 0.0870 \\[8pt]

\multicolumn{6}{l}{\textit{Panel E. Family-size filtering}} \\

Excluded by family-size filter
& $-0.0013$
& $-0.0098^{***}$
& $-0.0010^{***}$
& $-0.1452^{***}$
& $ 0.0199^{***}$ \\
& $(0.0013)$
& $(0.0005)$
& $(0.0002)$
& $(0.0110)$
& $(0.0007)$ \\

Observations
& 2,924,122
& 2,327,758
& 2,924,122
& 2,924,122
& 2,053,054 \\

Pseudo $R^2$ / $R^2$
& 0.0284
& 0.1258
& 0.0203
& 0.1099
& 0.0873 \\[8pt]

\multicolumn{6}{l}{\textit{Panel F. Citation filtering}} \\

Excluded by citation filter
& $-0.0770^{***}$
& $-0.0147^{***}$
& $ 0.0001$
& --
& -- \\
& $(0.0008)$
& $(0.0004)$
& $(0.0001)$
& 
&  \\

Observations
& 2,924,122
& 2,327,758
& 2,924,122
& 
&  \\

Pseudo $R^2$ / $R^2$
& 0.0290
& 0.1262
& 0.0202
& 
&  \\[8pt]

\multicolumn{6}{l}{\textit{Panel G. Triadic filtering}} \\

Excluded by triadic filter
& $ 0.0406^{***}$
& $ 0.0195^{***}$
& $-0.0006^{**}$
& $ 0.0210^{*}$
& $-0.0000$ \\
& $(0.0014)$
& $(0.0005)$
& $(0.0003)$
& $(0.0108)$
& $(0.0007)$ \\

Observations
& 2,924,122
& 2,327,758
& 2,924,122
& 2,924,122
& 2,053,054 \\

Pseudo $R^2$ / $R^2$
& 0.0285
& 0.1261
& 0.0202
& 0.1099
& 0.0870 \\

\bottomrule
\end{tabular}

\begin{minipage}{0.97\textwidth}
\vspace{2pt}
\footnotesize
\textit{Notes:} Robust standard errors are reported in parentheses. Each
coefficient refers to an indicator identifying patents affected by, or excluded
under, the corresponding mechanism. All regressions include 8-digit Y02
technology controls, patent office controls, and five-year period fixed effects.
The Scope models are estimated using Poisson regressions. The remaining models are estimated using OLS. Novelty is measured at the four-digit level. The citation-filter indicator is omitted from the generality model because of collinearity. $^{***}p<0.01$, $^{**}p<0.05$, $^{*}p<0.10$.
\end{minipage}
\end{table}

Table~\ref{tab:excluded_patent_characteristics} reports the differences in patent characteristics between families affected by each mechanism and their respective comparison groups, as defined in Section~\ref{subsec:patent_characteristics_methods}. 
The estimates provide evidence that changes in the observed boundary of green technology are associated with systematic differences in the nature of the inventions excluded by the different mechanisms. Patents affected by the reclassification and set expansion mechanisms exhibit slightly lower technological complexity, as shown by the patent scope and originality indicators relative to other green patent families included in the sample. At the same time, they display higher novelty. This pattern suggests that the inventions incorporated through retrospective classification changes or through the expansion of classification coverage tend to draw on narrower and less diverse knowledge bases, while being more likely to introduce technological combinations that had not previously appeared in the relevant domain. 

The results for technological impact reveal a distinctive pattern. Patents affected by reclassification and set expansion receive more forward citations, suggesting that they generate comparatively stronger subsequent technological influence. This is particularly important for assessing the broader spillovers within the green technological domain. Excluding these patents from the observed green domain may therefore underestimate the extent to which green inventions contribute to follow-on technological development. At the same time, their lower generality indicates that these spillovers may be concentrated within a narrower set of technological fields. 

The effects of filtering are substantially more heterogeneous. Excluding patents that lack an EP filing disproportionately removes inventions with lower scope, originality, and novelty. These patents, nevertheless, receive slightly more citations and display marginally higher generality. The US-based filter is considerably more selective: the excluded patents are characterised by markedly lower complexity and novelty and receive fewer citations, although the difference in generality is not statistically significant.

The family size filter also changes the composition of the observed patent population, although its effects vary across indicators. Patents excluded by this criterion do not differ significantly in scope, but have lower originality, novelty, and citation impact. At the same time, they exhibit higher generality.

As for the forward citation filter, the table reports only the coefficients for the ex-ante indicators (i.e., scope, originality, and novelty). Indeed, for impact indicators the result is mechanical because citation performance directly defines the selection criterion. However, the citation filter also excludes patents with lower scope and originality, while producing no statistically significant difference in novelty.

The triadic filter produces a different and, in some respects, opposite pattern. Patents excluded because they are not triadic have greater scope and originality, but slightly lower novelty. On the other hand, they are characterized by more impact on subsequent inventions.

These findings show that boundary mechanisms may select specific types of green invention. Reclassification and set expansion reveal a relatively consistent profile of more novel inventions, with stronger yet more technologically concentrated citation impact. Filtering mechanisms, by contrast, generate highly criterion-specific forms of selection.



\section{Discussion and implications}
\label{sec:discussion}
The results show that reclassification, set expansion, and filtering shape the observed boundary of green technology in distinct ways. Reclassification primarily alters the technological composition, set expansion affects geographical visibility, and filtering changes both the size and composition of the analytical sample. Together, these mechanisms influence assessments of technological opportunities, leadership, and the nature of green inventive activity.

The reclassification mechanism has its largest effect on the structure of the technological space. It is concentrated in specific domains, particularly energy storage, low-carbon production, renewable energy, and other rapidly developing or system-level technologies. The principal impact is therefore on the technological boundaries within which capabilities and opportunities are measured. This result is consistent with the information infrastructure literature, which emphasizes that categories do not simply describe reality but stabilize gradually as knowledge and institutional judgments evolve \citep{bowker2000sorting, mohlin2014optimal}. Different classification vintages may therefore produce different representations of the same underlying inventive activity, especially in emerging or contested domains whose environmental relevance becomes clearer only over time \citep{nelson2014innovation, lafond2019long}. In this sense, reclassification operates primarily as a mechanism of delayed technological recognition.

The comparatively limited geographical effect of reclassification suggests that it is less likely to radically alter assessments of country leadership or accumulated national capabilities. 
Nevertheless, countries specialized in subsequently reclassified fields may appear to strengthen their green technological position even when the underlying stock of inventions remains unchanged.
Hence, observed changes in specialization should not automatically be interpreted as changes in capabilities. They may instead reflect a change in the classificatory boundary through which those capabilities are made visible. This is relevant to the literature on related diversification and technological opportunity spaces, which generally treats observed patent classes as stable units from which relatedness and specialization can be inferred \citep{neffke2011regions, boschma2017relatedness, montresor2020green}. Our findings indicate that the opportunity space is partly endogenous to the evolution of the classification infrastructure itself. 
The characteristics of reclassified patents reinforce this interpretation. 
Reclassified patents appear relatively specialized and technologically influential.

Set expansion primarily affects the geography of observed green technology because newly visible families are disproportionately associated with Japanese, Chinese, and Korean patent offices. Earlier database releases may therefore understate the contributions and technological specializations of these jurisdictions, causing measured catching-up to combine changes in inventive capabilities with changes in institutional visibility \citep{lee2017catch, kwon2017international, corrocher2024green}.

Moreover, the set expansion mechanism also reshapes the technological distribution of green innovation, reflecting the fact that the newly covered jurisdictions differ in their industrial structures, knowledge bases, and patterns of specialization. Expanding coverage of specialized countries therefore changes not only the geographical distribution of patents but also the composition of the observed technological landscape. This result is consistent with the literature on the geography of transitions, which stresses that technological trajectories emerge unevenly across places because capabilities, institutions, markets, and policy priorities are spatially differentiated \citep{coenen2012geography, truffer2015geography}. 

Newly visible patents are more specialized and highly cited, but their influence extends across a narrower range of technological fields. This is important for studies of green knowledge spillovers and the double externality associated with environmental innovation \citep{jaffe2005tale}. If influential patents from late-covered offices are missing, estimates of the composition and geographical distribution of technological impact within the green patent population may be sensitive to database coverage. This issue is conceptually distinct from comparisons between green and non-green inventions \citep{dechezlepretre2014knowledge, barbieri2020knowledge, popp2012does}.

Filtering, by contrast, has its largest effect on the size of the observed population and also alters its geographical composition, while leaving the broad ranking of technological domains comparatively stable. Most filters substantially reduce the number of observable green patents, but the ranking of major technological domains remains comparatively stable. Family size, office-based, and triadic family filters disproportionately reduce the representation of patents associated with Asian offices. Office-based and international-family filters tend to reproduce the perspective of historically central patenting systems. They may underestimate the presence of locally oriented inventions, domestic-market specialists, and firms whose patenting strategies do not involve the EPO, USPTO, or multiple international jurisdictions. Finally, the citation filter leads to a lower geographic distortion, but remains sensitive to citation windows, patent office practices, and examiner behavior.
The effects on invention characteristics vary across filters, showing that they may select different profiles of green inventions. Depending on the criterion, filtering can select heterogeneous types of green inventions.

\subsection{Implications for research on green technological change}
\label{subsec:discussion_research}

Our findings have implications for research on directed technological change and the drivers of green innovation. Studies commonly use green patent counts to estimate responses to environmental regulation, carbon prices, subsidies, market demand, and public R\&D \citep{popp2002induced, popp2019complete, dugoua2023induced, acemoglu2023climate}. We show that the dependent variable in such analyses is sensitive to classification vintage, institutional coverage, and filtering choices. 
Estimated inducement effects may consequently be easier to detect in established and well-covered technological domains than in emerging fields whose climate relevance or institutional association becomes visible with delay. These choices may also understate the knowledge impact of newly visible green inventions or overrepresent spillovers within established international patenting networks.

The results also qualify patent-based analyses of technological relatedness and diversification. Such studies infer opportunity spaces from the observed distribution and proximity of technological capabilities \citep{neffke2011regions, boschma2017relatedness, montresor2020green, santoalha2021diversifying}. Our findings show that this empirical map depends partly on the information infrastructure used to construct it. Earlier classification vintages may underrepresent emerging trajectories, while incomplete jurisdictional coverage and restrictive filters may remove locally oriented or weakly internationalized capabilities. 

These findings do not reduce the value of classification-based indicators, but they identify information infrastructure as an important dimension of patent analysis. Studies need to pay attention to these choices, distinguish changes in inventive activity from changes in observability, and test whether conclusions are robust to filtering criteria.

\subsection{Implications for policy}

Patent-based indicators are widely used to monitor technological progress, benchmark countries, evaluate climate and innovation policies, and identify fields for public support \citep{hascic2015measuring, popp2019complete}. Our analysis highlights that the information infrastructure that may be used to develop patent indicators is not a neutral representation of an independently observable technological landscape. Classificatory and institutional infrastructures through which green inventions become visible can influence which technologies appear legitimate, which countries appear technologically capable, and which trajectories attract policy attention.

A first implication concerns technology governance and path dependency. Classification infrastructures determine whether and when inventions are recognized as part of the green domain, which may be especially consequential for emerging, contested, and system-level technologies whose environmental relevance depends on broader socio-technical configurations \citep{geels2002technological, markard2012sustainability}. Technologies that fit established categories are more readily incorporated into policy assessments, whereas less stabilized trajectories may remain comparatively invisible. Because such indicators inform expectations, strategic planning, and resource allocation, differences in visibility may reinforce attention to already recognized technological pathways. Our analysis does not establish these policy effects directly, but it identifies the measurement mechanism through which they may arise.

A second implication concerns measurement-induced policy errors. Evaluations of carbon pricing, environmental regulation, innovation subsidies, public R\&D, standards, and mission-oriented programmes often rely on green patent indicators as measures of policy-induced inventive activity \citep{popp2002induced, popp2019complete, dugoua2023induced, acemoglu2023climate}. However, an apparent increase in green patenting may partly reflect changes in classification or database coverage rather than a contemporaneous response to policy. 

These measurement problems may distort international and regional performance benchmarks, overstate the position of incumbent technological leaders, and understate catching-up in less visible jurisdictions. They may also bias the allocation of innovation subsidies and public R\&D toward well-established technological fields, while overlooking emerging opportunities and transition bottlenecks. In addition, restrictive patent filters may underrepresent locally oriented inventions and firms that do not seek protection through established international filing routes. Policy conclusions based on such indicators may therefore favour internationally embedded actors and mature technologies even when other inventions make relevant contributions to the transition.

The problem extends to the assessment of technological quality and spillovers. Patent filters based on citations, family size, or filings at selected offices embody different concepts of technological or economic relevance. Their use can consequently change which inventions are treated as valuable and which forms of technological impact enter the evidence base for policy. A policy designed around a single indicator may therefore support a narrow profile of innovation rather than the broader range of inventions needed for a systemic transition.

A third implication concerns the institutional infrastructure of innovation measurement. Uneven CPC coverage may reflect differences in patent office classification capacity, database interoperability, retrospective coverage, and participation in international classification systems. Improving comparability therefore requires more transparent reporting of jurisdictional coverage, stronger interoperability, and broader participation in category revision. 

\subsection{Implications for managers and technology strategy}
\label{subsec:discussion_strategy}

Firms, investors, and corporate decision makers use patent data to assess technological positions, monitor competitors, identify opportunities, and guide R\&D and intellectual-property strategies \citep{jaffe1989characterizing, jaffe2017citations, higham2021patent}. However, analyses based on a single classification vintage or filtering criterion may distort the competitive landscape by underrepresenting newly classified fields, late-covered jurisdictions, or inventions associated with less internationalized filing strategies. Technology intelligence should consequently complement Y02 searches with neighboring patent classes, keyword searches, scientific publications, standards, and early market developments \citep{grodal2015coevolution, kovacs2021categories, pontikes2012two}.

Database coverage and indicator choice are equally important for competitor benchmarking. Older PATSTAT editions or restrictive filters, such as triadic families and large international families, may understate the capabilities of firms and countries associated with gradually covered patent offices, domestic-market specialists, and locally oriented applicants, particularly in Asian latecomer economies \citep{kwon2017international, corrocher2024green}. 

\subsection{Limitations and future research}
\label{subsec:limitations}

Several limitations suggest directions for future research. The analysis considers only two PATSTAT releases and the CPC Y02 system, so additional database vintages, classification schemes, and technological domains are needed to determine whether reclassification and coverage effects are specific to green technologies or more general features of rapidly evolving fields. Because alternative taxonomies and text-based methods identify only partially overlapping sets of green inventions, future work should also examine how institutionalized classifications interact with researcher-constructed and keyword-based approaches \citep{hascic2015measuring,favot2023green,daly2019mining}. Moreover, the indicators used here do not reveal whether newly observed inventions reinforce existing technological trajectories or create new ones; combining longitudinal citation and recombination measures with qualitative evidence could help distinguish path-continuing from path-creating change and assess whether estimates of policy inducement vary across database and classification vintages \citep{dosi1982technological, nelson1982evolutionary, hotte2023demand}. Finally, institutional research should investigate how patent offices, firms, classification specialists, and policy organizations shape and revise green-technology categories, since the boundaries of green technology are not only conceptually defined but also embedded in the information systems used to observe technological change.

\section*{Acknowledgments}
The present paper has benefited from feedback received at the Oxford INET seminar; Italian Association of Environmental and Resource Economists (IAERE); Eco-Innovation Society workshop at HEC Paris; Kedge Business School research seminar; the Munich Summer Institute and Annual Workshop on Economics with Heterogeneous Interacting Agents (WEHIA). Peter would like to thank the OMPTEC programme at the Oxford Martin School for making this research possible.

\printbibliography

\newpage
\renewcommand{\appendixname}{Appendix}
\renewcommand{\thesection}{\Alph{section}} \setcounter{section}{0}
\renewcommand{\thefigure}{\Alph{section}.\arabic{figure}} \setcounter{figure}{0}
\renewcommand{\thetable}{\Alph{section}.\arabic{table}} \setcounter{table}{0}
\renewcommand{\theequation}{\Alph{section}.\arabic{table}} \setcounter{equation}{0}
\part*{Appendix}
\appendix
\FloatBarrier
\section{Appendices}

\subsection{Changes in the Y02 scheme over time}
\label{app:y02}

Table~\ref{tab:history_recla} summarizes the main scheme changes between 2015 and 2020.

\begin{table}[ht]
    \centering
    \footnotesize
    \caption{Changes to the Y02 scheme, 2015--2020} \resizebox{\textwidth}{!}{%
    \begin{tabular}{lccccc}
\toprule
   Date & CPC subclass & Deleted symbols & New symbols & Title changes & Indentation changes \\ 
\midrule
2020/08 &          Y02A &             X &        &              X &                   X \\
2020/08 &         Y02B &             X &        &              X &                   X \\
2020/08 &         Y02C &             X &         X &              X &                  \\
2020/08 &         Y02D &             X &         X &              X &                  \\
2020/08 &         Y02E &             X &        &              X &                   X \\
2020/08 &         Y02P &             X &        &              X &                   X \\
2020/08 &         Y02T &             X &        &              X &                   X \\
2020/08 &         Y02W &             X &        &              X &                   X \\ \midrule
2019/08 &         Y02C &            &        &              X &                  \\ \midrule
2018/05 &         Y02B &             X &         X &              X &                  \\
2018/05 &         Y02T &             X &         X &              X &                  \\
2018/05 &         Y02E & &         X &              X &                  \\ \midrule
2018/01 &         Y02A &            &         X &             &                  \\
2018/01 &         Y02B &             X &        &             &                  \\
2018/01 &         Y02D &            &         X &             &   \\
\midrule
2015/11 &         Y02T &             X &        &             &                  \\
2015/11 &         Y02P &            &         X &             &                  \\ \midrule
2015/05 &         Y02W &            &         X &             &                  \\ 
\bottomrule
\end{tabular}
}
\label{tab:history_recla}

\noindent \justifying

Notes: An X indicates that the official CPC Compilation of Changes for the stated release records at least one corresponding change within the subclass. Column 2 indicates the CPC subclass affected by a change; columns 3 and 4 indicate whether groups have been deleted or newly introduced to these subclasses; column 5 indicates whether the description of the technology subclass has been amended; and column 6 indicates changes in the indentation, i.e., the hierarchical level of the code. Source: Authors' compilation based on the official EPO--USPTO CPC archive and Compilation of Changes files \citep{epo_uspto_cpc_archive, epo_uspto_cpc_compilation_spec}.
\end{table} 

\subsection{Relation between class size and reclassification magnitude}
\label{app:reclass_class}

This appendix investigates the general relation between class size (in number of patents) and the number of reclassifications happening within those classes. 

\begin{figure}[H]
\centering
\includegraphics[width=0.75\linewidth]{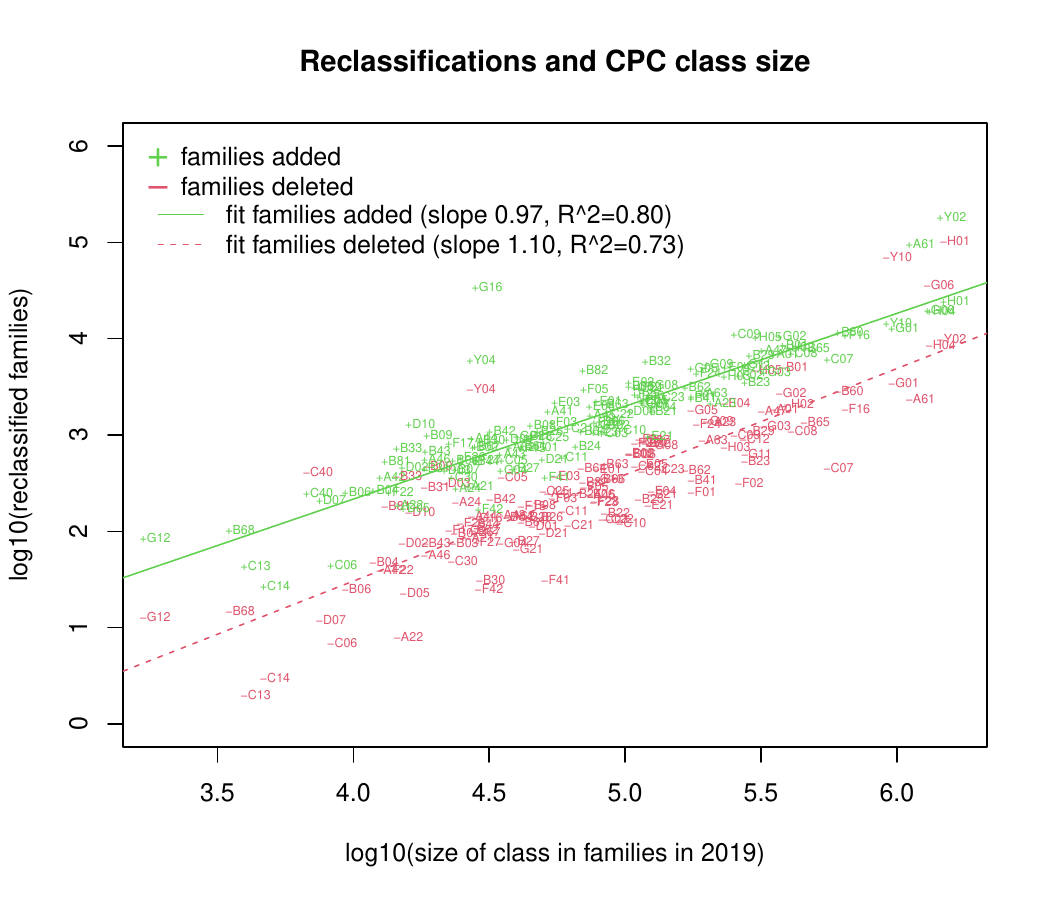}
\caption{Reclassification and class size}
\label{subfig:recla_class_size_all}
    \justifying \footnotesize

This figure illustrates the relationship between the number of reclassifications (DOCDB patent families being added or deleted from a given class) and the size of the class, measured by the number of DOCDB patent families per class. We follow the same specifications as the main analysis in the text, i.e, the reclassifications are measured from the changes between the 2019 and 2023 versions of the CPC system, without imposing a quality filter, only selecting families with earliest filing year prior to 2017 and the classifications for each family are the pooled classifications of each of its family members. The slopes of the fits are close to 1, which suggests a direct proportionality between class size and reclassifications. Y02 is found at the top right. 

\end{figure}

We find that larger patent classes (measured by the number of patents in those classes) show larger number of positive and negative fluctuations, indicated by a high number of patents being added or deleted from a given class. In Figure ~\ref{subfig:recla_class_size_all} we plot the number of reclassifications for class size for all CPC classes on a log-log scale. The slopes of the fits are both close to 1, so even though we have a log-log scale, this suggests direct proportionality between class size and number of reclassifications. This relation suggests that, when researching the magnitude of reclassification of a class, it makes sense to consider the number of reclassifications relative to class size (or, the 'reclassification percentage').

Not all classes in Figure ~\ref{subfig:recla_class_size_all} are found neatly on the proportionality line: if a class is far above (below) the line, this indicates that even accounting for class size, there are relatively many (few) reclassifications in that class. Green technology (Y02) is on the top right, far above the fit. This means that green technology is reclassified relatively strongly not only in an absolute but also in a relative sense. 

\begin{figure}
    \centering
            \includegraphics[width=\linewidth]{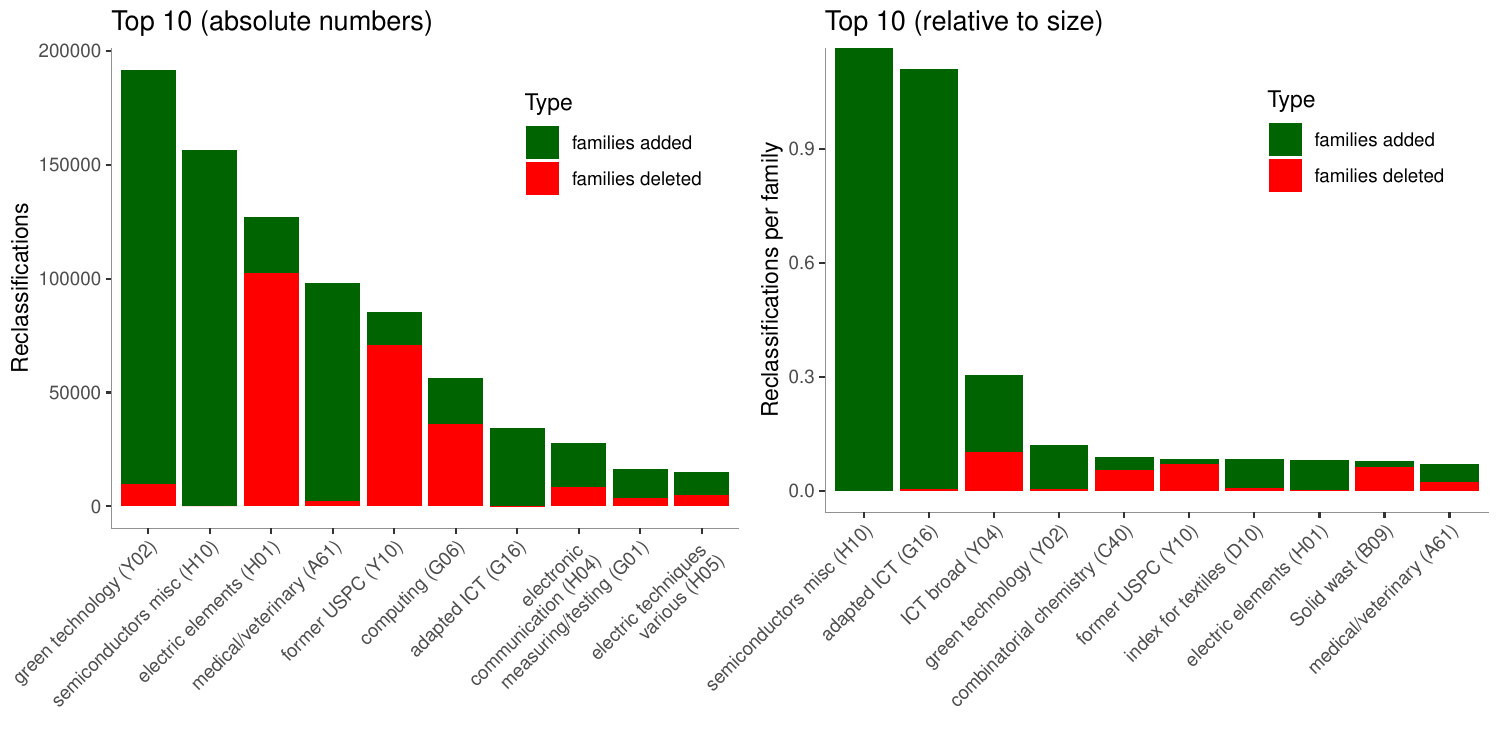}
        \caption{Patent classes with most frequent reclassifications}
    \label{fig:top10_recla_class}

    \justifying \footnotesize

    Notes: These figures show the top-10 most dynamically changing patent classes and subclasses, identified by the largest number of patents being removed or added to the class. On the left we have a ranking according to absolute changes (in number of reclassifications) on the right we have a ranking according to relative changes (in number of reclassifications per family). For these rankings, we selected only classes larger than a 1000 families.
\end{figure}

We also looked into those CPC classes and subclasses that are most dynamically changing by counts of patents being added and removed. Figure ~\ref{fig:top10_recla_class} shows the top-10 CPC classes ranked by their absolute (left panel) and relative (right panel) number of patents being migrated from and to the class (where we divide by the number of patents in those classes in 2023, we only consider classes larger than a 1000 families). Considering the absolute class rankings first, we first observe that the reclassifications are distributed rather unequally, with only a small number of 'turbulent' classes being reclassified very frequently. Again we observe that green technology is one of these classes. Moreover, it is interesting that where the reclassifications of green technology mostly consist of patents added, the reclassification of Electrical Elements (ranked second) consists of many more patents removed. Learning from the earlier mentioned relation between reclassification and class size, we also consider the class ranking relative to size in Figure  ~\ref{fig:top10_recla_class}. The first two classes can be considered outliers: the first is a miscellaneous class for semiconductors introduced between 2019 and 2023 and the second, 'adapted ICT', was also introduced shortly before 2019. Naturally, newly introduced classes show a disproportionately large reclassification fraction. Again green technology scores high, with only three other classes showing a greater reclassification fraction. Finally, we note that green technology is the only class with a high rank both in an absolute as well as in a relative sense.  

\subsection{CPC coverage of Chinese, Korean and Japanese patent documents between 2015-2026}\label{app:cpc_coverage}

In this appendix, we include the data we could find on CPC coverage of Chinese (Table \ref{tab:cpc_coverage_china}), Korean (Table \ref{tab:cpc_coverage_korea}) and Japanese (Table \ref{tab:cpc_coverage_japan}) patent documents (including both applications and related legal documents) between 2015 and 2026. The numbers are taken mostly from the CPC year reports, to which a separate reference list is included below. Despite the fact that we could not find data for all years, a gradual increase in coverage can be observed at least for the Chinese and Korean patents. Still, it is clear that in the 2010s, the CPC coverage of all three patent offices was still limited.    

\begin{table}[htbp]
\centering
\small
\caption{CPC coverage of Chinese patent documents/applications}
\label{tab:cpc_coverage_china}
\begin{tabular}{@{}llrrr@{}}
\toprule
\textbf{Date} & \textbf{Source} & \textbf{Total docs/apps} & \textbf{CPC classified} & \textbf{Coverage} \\
\midrule
Mar. 2015 & R2015 & 8\,579\,224  & 1\,627\,479  & 18.8\% \\
Feb. 2017 & R2017 & 12\,159\,983 & 2\,406\,586  & 19.8\% \\
Dec. 2018 & R2018 & 17\,286\,037 & 5\,473\,797  & 31.7\% \\
Feb. 2020 & R2020 & 20\,632\,269 & 6\,903\,470  & 33.5\% \\
Jan. 2022 & R2022 & --           & 7\,426\,664$^{a}$  & -- \\
Jan. 2023 & R2023 & 33\,630\,811 & 12\,595\,182 & 37.5\% \\
Jan. 2024 & R2024 & --           & 14\,818\,679$^{b}$ & -- \\
Feb. 2025 & R2025 & 41\,798\,611$^{c}$ & -- & -- \\
Feb. 2026 & R2026 & 45\,293\,421 & 19\,146\,659 & 42.3\% \\
\bottomrule
\end{tabular}

\vspace{0.15cm}
\footnotesize
$^{a}$ The 2022 deck reports only applications classified by the CPC office/national office. \\
$^{b}$ The 2024 deck reports publications classified in CPC, but no total denominator. \\
$^{c}$ The 2025 deck gives a country-level application count and a national-office CPC count, but no separate overall CPC-classified denominator; coverage is therefore not calculated.
\end{table}

\begin{table}[htbp]
\centering
\small
\caption{CPC coverage of Korean patent documents/applications}
\label{tab:cpc_coverage_korea}
\begin{tabular}{@{}llrrr@{}}
\toprule
\textbf{Date} & \textbf{Source} & \textbf{Total docs/apps} & \textbf{CPC classified} & \textbf{Coverage} \\
\midrule
Mar. 2015 & R2015 & 2\,810\,926 & 878\,787   & 31.3\% \\
Feb. 2017 & R2017 & 3\,245\,572 & 1\,420\,956 & 43.8\% \\
Dec. 2018 & R2018 & 3\,705\,485 & 2\,482\,712 & 67.0\% \\
Feb. 2020 & R2020 & 4\,153\,209 & 2\,748\,463 & 66.2\% \\
Jan. 2022 & R2022 & --          & 3\,873\,874$^{a}$ & -- \\
Jan. 2023 & R2023 & 4\,886\,339 & 4\,279\,287 & 87.6\% \\
Jan. 2024 & R2024 & --          & 4\,467\,302$^{b}$ & -- \\
Feb. 2025 & R2025 & 5\,315\,121$^{c}$ & -- & -- \\
Feb. 2026 & R2026 & 5\,525\,162 & 4\,869\,576 & 88.1\% \\
\bottomrule
\end{tabular}

\vspace{0.15cm}
\footnotesize
$^{a}$ The 2022 deck reports only applications classified by the CPC office/national office. \\
$^{b}$ The 2024 deck reports publications classified in CPC, but no total denominator. \\
$^{c}$ The 2025 deck gives a country-level application count and a national-office CPC count, but no separate overall CPC-classified denominator; coverage is therefore not calculated.
\end{table}

\begin{table}[htbp]
\centering
\small
\caption{CPC coverage of Japanese patent documents}
\label{tab:cpc_coverage_japan}
\begin{tabular}{@{}llrrr@{}}
\toprule
\textbf{Date} & \textbf{Source} & \textbf{Total docs/apps} & \textbf{CPC classified} & \textbf{Coverage} \\
\midrule
15 Mar. 2015 & R2015 & 16\,886\,236 & 4\,123\,806 & 24.4\% \\
21 Dec. 2018 & R2018 & 18\,179\,959 & 4\,848\,814 & 26.7\% \\
\bottomrule
\end{tabular}

\vspace{0.15cm}
\begin{minipage}{0.92\linewidth}
\footnotesize
Notes: The 2015 row refers to documents present in DocDB and classified in CPC through direct classification and family propagation. The 2018 row refers to documents/publications classified in CPC at document or family level. Rows for 2020--2026 are not added because the CPC update decks inspected do not report Japan in the same denominator/numerator format.
\end{minipage}
\end{table}

The list of sources made to create these tables are 

\begin{itemize}

\item[R2015] EPO and USPTO. 2015. \emph{2nd CPC Annual Meeting: EPO--USPTO Presentation -- CPC Status Update}. Presentation by Nigel Lombard, 1 May 2015. Used for the CPC coverage table with status date 15 March 2015.
\url{https://link.epo.org/cpc/EPO-USPTO_presentation_Lombard_May2015.pdf}

\item[R2016] EPO and USPTO. 2016. \emph{3rd CPC Annual Meeting with National Offices Classifying in CPC}. CPC annual meeting presentation, status date 13 September 2016. Used for the 2016 China and Korea CPC coverage rows.
\url{https://www.cooperativepatentclassification.org/sites/default/files/attachments/618ccb71-8276-43b8-85d3-434e09106470/CPC%2Bannual%2Bmeeting%2Bpresentation%2Bfinal.pdf}

\item[R2017] EPO and USPTO. 2017. \emph{Recent Developments in the Cooperative Patent Classification}. Presentation for the WIPO Fourth IPC Workshop, Geneva, 21 February 2017. Used for the CPC coverage table with status date 1 February 2017.
\url{https://www.wipo.int/edocs/mdocs/classifications/en/ipc_wk_ge_17/ipc_wk_ge_17_item2_4_epo_uspto.pdf}

\item[R2018a] EPO and USPTO. 2018. \emph{5th CPC Annual Meeting with National Offices: CPC Status Update}. Geneva, 6 February 2018. Used for the early-2018 China and Korea CPC coverage rows.
\url{https://link.epo.org/cpc/CPC%2BAnnual%2BMeeting%2B2018_USPTO%2BFINAL%2BFebruary_2018%2BFINAL%2BFINAL.pdf}

\item[R2018b] EPO and USPTO. 2019. \emph{Cooperative Patent Classification Annual Report 2017/2018}. Table 3.1, ``Documents classified in CPC at document or family level'', status date 21 December 2018. Used for China, Korea and Japan.
\url{https://link.epo.org/cpc/CPCAnnualReport_published.pdf}

\item[R2020] EPO and USPTO. 2020. \emph{7th CPC Annual Meeting with National Offices: CPC Update}. Geneva, 18 February 2020. Used for the China and Korea rows with status date 5 February 2020.
\url{https://link.epo.org/cpc/7th_cpc_annual_meeting_nos_cpc_update.pdf}

\item[R2022] EPO and USPTO. 2022. \emph{9th CPC Annual Meeting with National Offices: CPC Update}. Geneva, 15--16 March 2022. Used for the China and Korea CPC-office classification rows with status date 3 January 2022.
\url{https://link.epo.org/cpc/NOTES_VERSION_031022_9th_Annual_meeting_NOs}

\item[R2023] EPO and USPTO. 2023. \emph{10th CPC Annual Meeting with National Offices: CPC 10 Years' Update}. Geneva, 8--9 March 2023. Used for the China and Korea rows with status date 1 January 2023.
\url{https://link.epo.org/cpc/10th_CPC_Annual_meeting_with_NOs}

\item[R2024] EPO and USPTO. 2024. \emph{11th CPC Annual Meeting with Offices: Classifying in the CPC}. Geneva, 14 March 2024. Used for the China and Korea CPC-classified publication rows with status date 1 January 2024.
\url{https://link.epo.org/cpc/11th_CPC_Annual_meeting_NOs/CPC_update}

\item[R2025] EPO and USPTO. 2025. \emph{12th CPC Annual Meeting with Offices: Classifying in the CPC}. Geneva, 24 February 2025. Used for the 2025 CPC data rows.
\url{https://link.epo.org/cpc/12th_CPC_Annual_meeting/CPC_update}

\item[R2026] EPO and USPTO. 2026. \emph{13th CPC Annual Meeting with Offices: Classifying in the CPC}. Geneva, 23 February 2026. Used for the China and Korea rows with status date 1 February 2026.
\url{https://link.epo.org/cpc/13th_CPC_Annual_meeting/CPC_update}

\end{itemize}




\end{document}